\documentclass[12pt]{article}
 \usepackage[dvips]{graphicx}

\begin{document}

 \sffamily

 \title{You Can't Get Through Szekeres Wormholes ~~or \\
        Regularity, Topology and Causality in \\
        Quasi-Spherical Szekeres Models}

 \author{
 Charles Hellaby
 \thanks{This research was supported by a grant from the South African
         National Research Foundation,} \\
 Department of Mathematics and Applied Mathematics, \\
 University of Cape Town, Rondebosch 7701, South Africa \\
 {\tt cwh@maths.uct.ac.za}
 \and
 ~~~~~~~~~~~~~~~~~~~~and~~~~~~~~~~~~~~~~~~~~
 \and
 Andrzej Krasi\'{n}ski
 \thanks{and by the Polish Research Committee
         grant no 2 P03B 060 17.} \\
 N. Copernicus Astronomical Center, Polish Academy of Sciences, \\
 Bartycka 18, 00 716 Warszawa, Poland \\
 {\tt akr@camk.edu.pl}
 }

 \date{}

 \maketitle

 \begin{abstract}

The spherically symmetric dust model of Lema\^{\i}tre-Tolman can describe
wormholes, but the causal communication between the two asymptotic regions
through the neck is even less than in the vacuum
(Schwarzschild-Kruskal-Szekeres) case. We investigate the anisotropic
generalisation of the wormhole topology in the Szekeres model.  The function
$E(r, p, q)$ describes the deviation from spherical symmetry if $\partial_r E
\neq 0$, but this requires the mass to be increasing with radius, $\partial_r M
> 0$, i.e. non-zero density. We investigate the geometrical relations between
the mass dipole and the locii of apparent horizon and of shell-crossings. We
present the various conditions that ensure physically reasonable
quasi-spherical models, including a regular origin, regular maxima and minima
in the spatial sections, and the absence of shell-crossings. We show that
physically reasonable values of $\partial_r E \neq 0$ cannot compensate for the
effects of $\partial_r M > 0$ in any direction, so that communication through
the neck is still worse than in the vacuum.

 We also show that a handle topology cannot be created by identifying
hypersufaces in the two asymptotic regions on either side of a wormhole,
unless a surface layer is allowed at the junction.  This impossibility
includes the
 Schwarzschild-Kruskal-Szekeres case.
 \end{abstract}

 \begin{center}

 {PACS: \\
  04.20.Gz, spacetime topology \& causal structure \\
  04.40.Nr, spacetimes with fluids or fields \\
  04.70.Bw, classical black holes
 }
 \\[1mm]

 \\[1mm]

 {Short title:~~
 Wormholes and Other Quasi-Spherical Szekeres Models
 }
 \\[1mm]

 Submitted to:~~Physical Review D, June 2002
 \\[1mm]

 {\tt gr-qc/0206052}
 \\[1mm]

 \end{center}

 \section{Introduction}

 The Szekeres metric is a dust model, which has no Killing vectors
\cite{BoSuTo77}, but contains the
 Lema\^{\i}tre-Tolman (LT) model as the spherically symmetric special
case, which itself contains the Schwarzschild-Kruskal-Szekeres \cite{Kru60,
Sze60} manifold as the vacuum case.  As with the LT model, it is written in
synchronous coordinates, and the particles of dust are comoving.  The constant
time slices are foliated by
 2-surfaces of constant coordinate $r$, which have
 2-metrics of spheres, planes or
 pseudo-spheres, depending on the value of parameter $\epsilon$.  See
\cite{Kra97} for a review of its known properties.

     Despite the inhomogeneity of the model, and the lack of Killing
vectors, any surface of constant coordinate `radius' $r$ in the $\epsilon
= +1$ case can be matched onto a Schwarzschild vacuum metric \cite{Bon76a,
Bon76b}, and any surface of constant time $t$ is conformally flat
\cite{BeEaOl77}.

 We here investigate the topological and causal properties of the
quasi-spherical case, $\epsilon = +1$, subject to the requirements for a
physically reasonable model. Reasonability requirements include, well behaved
metric components, non-divergent density and curvature, regular spherical
origins, regular maxima and minima in the spatial sections, and prohibition of
shell crossings. Choosing well behaved coordinates also assists in avoiding the
confusion of coordinate singularities.

 Studying such models of low symmetry is important, so that one can check
which properties of spherically symmetric investigations of cosmology and
gravitational collapse are preserved, and which are not.

The subjects studied in this paper in some detail are the following:

1. The dipole-like variation of mass-density; the locus of its poles and of the
equator, and the images of the equator under the Riemann projection.

2. Conditions for regularity of the geometry at the origin $R = 0$.

3. Intersections of the shell crossings with the surfaces of constant $(t, r)$,
and conditions for avoidance of shell crossings.

4. Conditions for regular maxima and minima at necks and bellies.

5. Conditions for a handle topology of a $t =$ const space, and the
impossibility of preserving this topology during evolution of the model.

6. Apparent horizons -- their shape, intersections with the surfaces of
constant $(t, r)$, relations between these intersections and those of shell
crossings, and with the dipole equator, location of an AH with respect to the
$R = 2M$ hypersurface, the intersection of an AH with a neck.

7. The impossibility of sending a light ray through the neck so that it would
emerge from under the AH on the other side.

8. Numerical examples of light paths traversing the neck and of those going in
its vicinity.

 \section{The Szekeres Metric}

     The LT-type Szekeres metric \cite{Sze75a} is\footnote{The results
presented in Ref. \cite{Sze75a} contain a few misleading typos that were
corrected in Ref. \cite{BoTo1976}. The notation used here does not follow the
traditional one}:
 \begin{equation}
  ds^2 = - dt^2 + \frac{(R' - R \frac{\textstyle E'}{\textstyle E})^2}
                       {(\epsilon + f)} dr^2
            + R^2 \frac{(dp^2 + dq^2)}{E^2} ,   \label{dsS}
 \end{equation}
 where ${}' \equiv \partial/\partial r$, $\epsilon = \pm1,0$ and $f = f(r)
\geq -\epsilon$ is an arbitrary function of $r$.

The function $E$ is given by
 \begin{equation}\label{Edeforig}
   E(r,p,q) = A (p^2 + q^2) + 2 B_1 p + 2 B_2 q + C ,
 \end{equation}
 where functions $A = A(r)$, $B_1 = B_1(r)$, $B_2 = B_2(r)$, and $C = C(r)$
satisfy the relation
 \begin{equation}\label{Econd}
   4(AC - B_1^2 - B_2^2) = \epsilon \;\;,\;\;\;\;\;\; \epsilon = 0, \pm 1,
 \end{equation}
but are otherwise arbitrary.

     The function $R = R(t,r)$ satisfies the Friedmann equation for dust
 \begin{equation}
   \dot{R}^2 = \frac{2M}{R} + f ,  \label{RdotSq}
 \end{equation}
 where $\dot{{}} \equiv \partial/\partial t$ and $M = M(r)$ is another
arbitrary function of coordinate ``radius", $r$.  It follows that the
acceleration of $R$ is always negative
 \begin{equation}
   \ddot{R} = \frac{-M}{R^2}.  \label{Rddot}
 \end{equation}
 Here $M(r)$ plays the role of an effective gravitational mass for
particles at comoving ``radius" $r$.  For $\epsilon = +1$, it is simply the
total gravitational mass within the sphere of radius $r$.  We assume $M \geq 0$
and $R \geq 0$.  In (\ref{RdotSq}) $f(r)$ represents twice the energy per unit
mass of the particles in the shells of matter at constant $r$, but in the
metric (\ref{dsS}) it also determines the geometry of the spatial sections $t =
$constant (c.f. \cite{Hel87}).  The evolution of $R$ depends on the value of
$f$; it can be: \\
 ${}$~~~hyperbolic, $f > 0$:
 \begin{eqnarray}
   R & = & \frac{M}{f} (\cosh \eta - 1),   \label{hypevRS}
   \\ \nonumber \\
   (\sinh \eta - \eta) & = & \frac{f^{3/2} \sigma (t - a)}{M},
   \label{hypevtS}
 \end{eqnarray}
 ${}$~~~parabolic, $f = 0$:
 \begin{eqnarray}
   R & = & M \frac{\eta^2}{2},   \label{parevRS}
   \\ \nonumber \\
   \frac{\eta^3}{6} & = & \frac{\sigma (t - a)}M,   \label{parevtS}
   \\ \nonumber \\
   \mbox{i.e.~~~~} R & = & \left( \frac{9 M (t - a)^2}{2} \right)
   ^{1/3},\label{evoRflat}
 \end{eqnarray}
 ${}$~~~elliptic, $f < 0$:
 \begin{eqnarray}
   R & = & \frac{M}{(-f)} (1 - \cos \eta),   \label{ellevRS}
   \\ \nonumber \\
   (\eta - \sin \eta) & = & \frac{(-f)^{3/2} \sigma (t - a)}{M},
   \label{ellevtS}
 \end{eqnarray}
 where $a = a(r)$ is the last arbitrary function, giving the local time of
the big bang or crunch $R = 0$ and $\sigma = \pm 1$ permits time reversal.
More correctly, the three types of evolution hold for $f/M^{2/3} >,=,< 0$,
since $f = 0$ at a spherical type origin for all 3 evolution types.  The
behaviour of $R(t,r)$ is identical to that in the LT model, and is
unaffected by $(p,q)$ variations.

  A more meaningful way to write $E$ is
 \begin{equation}
   E(r,p,q) = \frac{S}{2} \left\{ \left( \frac{p - P}{S} \right) ^2
             + \left( \frac{q - Q}{S} \right) ^2 + \epsilon \right\},
   \label{Edef}
 \end{equation}
 where $S = S(r)$, $P = P(r)$, and $Q = Q(r)$ are arbitrary functions,
and
 \begin{equation}
   A = \frac{1}{2S} ~~,~~~~~~ B_1 = \frac{-P}{2S} ~~,~~~~~~
   B_2 = \frac{-Q}{2S} ~~,~~~~~~ C = \frac{P^2 + Q^2 + \epsilon S^2}{2S}.
 \end{equation}

 The metric component
 \begin{equation}
   \frac{(dp^2 + dq^2)}{E^2}
 \end{equation}
 is actually the unit sphere, plane, pseudo-sphere in Riemann projection:
 \begin{eqnarray}
   \epsilon = +1~~~~~~~~~~
   \frac{(p - P)}{S} = \cot\left(\frac{\theta}{2}\right) \cos(\phi)
 ~~~~& , &~~~~
   \frac{(q - Q)}{S} = \cot\left(\frac{\theta}{2}\right) \sin(\phi),
   \label{Riemprojp} \\ \nonumber \\
   \epsilon = ~0~~~~~~~~~~~~~~
   \frac{(p - P)}{S} = \left(\frac{2}{\theta}\right) \cos(\phi)
 ~~~~& , &~~~~
   \frac{(q - Q)}{S} = \left(\frac{2}{\theta}\right) \sin(\phi),
   \label{Riemproj0} \\ \nonumber \\
   \epsilon = -1~~~~~~~~
   \frac{(p - P)}{S} = \coth\left(\frac{\theta}{2}\right) \cos(\phi)
 ~~~~& , &~~~~
   \frac{(q - Q)}{S} = \coth\left(\frac{\theta}{2}\right) \sin(\phi).
   \label{Riemprojm}
 \end{eqnarray}
 It seems reasonable to expect $S > 0$, but it is not obviously impossible
for $S$ to reach or pass through zero.

 \begin{figure}
 \begin{center}
 \parbox{14cm}{
 \includegraphics[scale = 0.9]{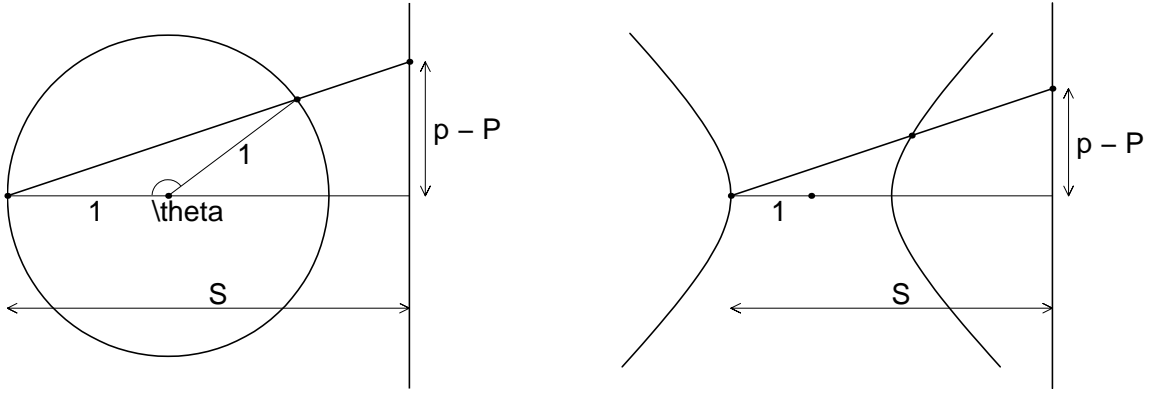}
 \caption{
 \label{Rieprofig}
 \footnotesize
 The Riemann projection from $(\theta, \phi)$ to $(p, q)$ coordinates for
spheres \& two-sheeted hyperboloids.  The diagrams show only the $\phi = 0,
\pi$ section, i.e. the $q = Q$ section.}
   }
 \end{center}
 \end{figure}

      The factor $\epsilon$ determines whether the
 $p$-$q$ 2-surfaces are spherical ($\epsilon = +1$),
 pseudo-spherical ($\epsilon = -1$), or planar ($\epsilon = 0$).  In other
words, it determines how the constant $r$
 2-surfaces foliate the 3-d spatial sections of constant $t$.  The
function $E$ determines how the coordinates $(p,q)$ map onto the unit
 2-sphere (plane,
 pseudo-sphere) at each value of $r$.  At each $r$ these
 2-surfaces are multiplied by the areal ``radius" $R = R(t,r)$ that
evolves with time.  Thus the
 $r$-$p$-$q$
 3-surfaces are constructed out of a sequence of
 2-dimensional spheres
 (pseudo-spheres, planes) that are not concentric, since the metric
component $g_{rr}$ depends on $p$ and $q$ as well as $r$ and $t$.

The $(p, q)$-coordinates in the cases $\epsilon = +1$ and $\epsilon = 0$ have
the range $(- \infty, +\infty)$. In the case $\epsilon = -1$, the
parametrization (\ref{Edef}) does not cover the subcases $A = 0$ and $C = 0$
(these subcases cannot occur with $\epsilon \geq 0$ because of (\ref{Econd})).
Coming back to (\ref{Edeforig}), we see that, for $\epsilon = -1$ and $A \neq
0$, $E$ is zero when
 $$
 \left(p + B_1/A\right)^2 + \left(q + B_2/A\right)^2 = 1/(4A)^2,
 $$
 $E$ is positive for $p$ and $q$ outside this circle, and is
negative for $p$ and $q$ inside it. Fig. \ref{Rieprofig} suggests that with
$\epsilon = -1$, we should rather take $(- E)$ as the metric function so that
$p$ and $q$ have finite rather than semi-infinite ranges. However, both the $E
> 0$ and $E < 0$ regions are Szekeres spacetimes because they are mapped one
onto another by
\begin{equation}\label{invers}
(p, q) = (p', q')/\left({p'}^2 + {q'}^2\right),
\end{equation}
the roles of $A$ and $C$ being interchanged after the transformation.

If $A \neq 0 = C$, then a nonzero $C$ is restored by a translation in the $(p,
q)$ plane. If $A = 0$, then the metric of the $(p, q)$-surface is brought back
to the standard Szekeres form with $A \neq 0 \neq C$ by a Haantjes
transformation (a conformal symmetry transformation of a flat space, see
\cite{Kra89} for a description) in the $(p, q)$ surface, which also restores
the appropriate form of $g_{rr}$.

The surface area of a $(t =$ const, $r = $const$)$ surface is finite only in
the $\epsilon = +1$ case, where it equals $4\pi R^2$. In the other two cases,
it is infinite.

     The 6 arbitrary functions $f$, $M$, $a$, $P$, $Q$ and $S$ represent 5
physical freedoms to control the inhomogeneity, plus a coordinate freedom 
to rescale $r$. 

     The density and Kretschmann scalar are functions of all four
coordinates%
 \footnote{
 There are only two independent curvature invariants in the Szekeres
metric, for which a good choice would be: $R = 8 \pi \rho$ and $C^{\alpha
\beta \gamma \delta} C_{\alpha \beta \gamma \delta} = 48 M^2 \{(R' - R M'
/ 3 M)^2\} / \{R^6 (R' - R E' / E)^2\} = (4/3) (8 \pi \overline{\rho}) (8
\pi \overline{\rho} - 8 \pi \rho)^2$
 --- a pure Ricci invariant and a pure Weyl invariant.  Though less tidy,
the $8 \pi \rho$ and $K$ used above will also suffice. We thank a referee for
pointing this out.
 }
 \begin{eqnarray}
   8 \pi \rho & = & G_{tt}
      = \frac{2 (M' - 3 M E' / E)}{R^2 (R' - R E' / E)},
         \label{RhoDef} \\ \nonumber \\
   {\cal K} & = &
      R^{\alpha \beta \gamma \delta} R_{\alpha \beta \gamma \delta}
      = (8 \pi)^2 \left[ \frac{4}{3} \overline{\rho}^2 - \frac{8}{3}
        \overline{\rho} \rho + 3 \rho^2 \right],
        \label{KretschDef}
 \end{eqnarray}
 where
 \begin{equation}
   8 \pi \overline{\rho} = \frac{6 M}{R^3}
 \end{equation}
 is the mean density within ``radius" $r$.  For all $\rho$ and
$\overline{\rho}$ we have ${\cal K} \geq 0$, but assumptions of positive
mass and density require $\rho \geq 0$ and $\overline{\rho} \geq 0$.
Clearly there are density and curvature singularities at $R = 0$
 --- the bang and/or crunch --- and at $R' = R E' / E$, $M' \neq 3 M E' /
E$
 --- shell crossings.  Additionally, $\rho$ but not ${\cal K}$ passes
through zero where $E'/E$ exceeds $M'/3M$.

 The matter flow $u^\alpha = \delta^\alpha_t$, with projection tensor
$h_{\alpha \beta} = g_{\alpha \beta} + u_\alpha u_\beta$, has the
following properties, which are almost trivial to calculate with GRTensor
\cite{MPL96}:
 \begin{eqnarray}
   \Theta ~= & \nabla_\alpha u^\alpha & =~
      \frac{(\dot{R}' - 3 \dot{R} E' / E + 2 R' \dot{R} / R)}
           {(R' - R E' / E)},
      \\ \nonumber \\
   a^\alpha ~= & u^\beta \, \nabla_\beta u^\alpha & =~ 0,
      \\ \nonumber \\
   \sigma^\alpha{}_\beta ~= &
      g^{\alpha \gamma} (\nabla_{(\gamma} u_{\alpha)}
      + u_{(\gamma} \, a_{\alpha)}) - \frac{\Theta}{3} \, h^\alpha_\beta
      & =~ \frac{(\dot{R}' - R' \dot{R} / R)}
      {3 (R' - R E' / E)}~ {\rm diag} (0, 2, -1, -1),
      \\ \nonumber \\
   \omega_{\alpha \beta} ~= & \nabla_{[\beta} u_{\alpha]}
      + u_{[\beta} \, a_{\alpha]} & =~ 0,
      \\ \nonumber \\
   {E^{\alpha}}_{\beta} ~= &
      {C^{\alpha}}_{\gamma \beta \delta} \, u^\gamma \, u^\delta
      & =~ \frac{M (R' - R M' / 3 M)}
      {R^3 (R' - R E' / E)}~ {\rm diag} (0, -2, 1, 1),
      \\ \nonumber \\
   H_{\alpha \beta} ~= & \frac{1}{2} \, \epsilon_{\alpha \gamma \mu \nu}
      \, C^{\mu \nu}{}_{\beta \delta} \, u^\gamma \, u^\delta & =~ 0.
 \end{eqnarray}

Note that the relation between the active gravitational mass $M$ and the
"sum-of-rest-masses" $\cal{M}$ is the same in the $\epsilon = +1$ Szekeres
model as in the LT model:

\begin{equation}\label{restMvsactM}
{\cal M}' = M'/\sqrt{1 + f}.
\end{equation}
The sum of the rest masses contained inside the sphere of coordinate radius $r$
at the time $t$ is defined by:

\begin{equation}\label{restMdef}
{\cal M} = \int \rho \sqrt{\left|g_3\right|} {\rm d}_3x,
\end{equation}
where $g_3$ is the determinant of the metric of the $t =$ const hypersurface,
and the integral is taken with respect to the variables $p$ and $q$ from $-
\infty$ to $+ \infty$, and with respect to $r$ from $r_0$ at the origin to the
current value $r$. We have

\begin{equation}\label{g3}
\sqrt{\left|g_3\right|} = \frac E{\sqrt{1 + f}} (R/E)^2(R/E)'.
\end{equation}
Consequently

\begin{equation}\label{restMeq}
{\cal M} = \frac 1{4\pi} \int_{- \infty}^{+\infty}{\rm d} q \int_{-
\infty}^{+\infty}{\rm d} p \int_{r_0}^{r}{\rm d} x \left[\frac E{\sqrt{1 +
f}}\left(M/E^3\right)'\right] (t, p, q, x).
\end{equation}
The term containing $E'$ is integrated by parts with respect to $x$ in order to
move the prime (which, in the integrand, means $\frac {\partial} {\partial x}$)
away from $E$ to functions that do not depend on $p$ and $q$. The result is

\begin{equation}\label{restMtransf}
{\cal M} = \frac 32 \left[\frac M{\sqrt{1 + f}} (r) - \frac M{\sqrt{1 + f}}
\left(r_0\right)\right] \cdot \frac 1 {4\pi} \int_{- \infty}^{+\infty}{\rm d} q
\int_{- \infty}^{+\infty}{\rm d} p E^{-2}
 $$ $$
 + \frac 1 {8\pi} \int_{-
\infty}^{+\infty}{\rm d} q \int_{- \infty}^{+\infty}{\rm d} p \int_{r_0}^{r}
{\rm d} x E^{-2}\left[\frac {3Mf'}{2(1 + f)^{3/2}} - \frac {M'}{\sqrt{1 +
f}}\right].
\end{equation}
We note that

\begin{equation}\label{unitsph}
\int_{- \infty}^{+\infty}{\rm d} q \int_{- \infty}^{+\infty}{\rm d} p E^{-2} =
4\pi
\end{equation}
(this is the surface area of a unit sphere), and so

\begin{equation}\label{finrestM}
{\cal M} = \frac 32 \left[\frac M{\sqrt{1 + f}} (r) - \frac M{\sqrt{1 + f}}
\left(r_0\right)\right] + \frac 12 \int_{r_0}^{r} \left[\frac {3Mf'}{2(1 +
f)^{3/2}} - \frac {M'}{\sqrt{1 + f}}\right]{\rm d} x.
\end{equation}
From here, we obtain the same relation that holds in the L--T model,
(\ref{restMvsactM}).

Note that this result holds only in the $\epsilon = +1$ Szekeres model (the
quasi-spherical one). With $\epsilon = 0$ or $\epsilon = -1$, the total surface
area of the $(p, q)$-surface is infinite, and so $\cal{M}$ cannot be defined.

 \subsection{Special Cases and Limits}\label{SCaL}

 The
 Lema\^{\i}tre-Tolman (LT) model is the spherically symmetric special case
$\epsilon = +1$, $E' = 0$.

     The vacuum case is $(M' - 3 M E' / E) = 0$ which gives $M \propto
E^3$, and this requires
 \begin{equation}
   M' = 0 = S' = P' = Q' = E'
 \end{equation}
 and any region over which this holds is the Schwarzschild metric in LT
coordinates \cite{Hel87}, with mass $M$.  (See \cite{Hel96b} for the full
transformation in the general case.)

 In the null limit, $f \rightarrow \infty$, in which the `dust' particles
move at light speed \cite{Hel96, Bon97}, the metric becomes a pure
radiation
 Robinson-Trautman metric of Petrov type D, as given in Exact Solutions
\cite{KSHM80}, equation (24.60) with (24.62)%
 \footnote{Ref \cite{Bon97} corrected ref \cite{Hel96}'s mistaken claim
that the null limit of Szekeres was a new metric.
 }
The Kinnersley rocket \cite{Kin69} is the $\epsilon = +1$ case of this null
limit, which is actually more general than the axially symmetric form given in
\cite{Hel96}.

The KS-type Szkeres metric was shown in \cite{Hel96} to be a special case of
the above LT-type metric, under a suitable limit.

 \subsection{Basic Physical Restrictions}\label{BPR}

 \begin{enumerate}

 \item   For a metric of Lorentzian signature $(-+++)$, we require
 \begin{equation}
   \epsilon + f \geq 0
 \end{equation}
 with equality only occuring where $(R' - R E' / E)^2 / (\epsilon + f) >
0$.  Clearly, pseudo-spherical foliations, $\epsilon = -1$, require $f
\geq 1$, and so are only possible for hyperbolic spatial sections, $f >
0$.  Similarly, planar foliations, $\epsilon = 0$, are only possible for
parabolic or hyperbolic spatial sections, $f \geq 0$, whereas spherical
foliations are possible for all $f \geq -1$.

 \item   We obviously choose the areal radius $R$ to be positive,
 \begin{equation}
   R \geq 0
 \end{equation}
 ($R = 0$ is either an origin, or the bang or crunch.  In no case is a
continuation to negative $R$ possible.)

 \item   The mass $M(r)$ must be positive, so that any vacuum exterior has
positive Schwarzschild mass,
 \begin{equation}
   M \geq 0.
 \end{equation}

 \item   We require the metric to be non degenerate \& non singular,
except at the bang or crunch.  Since $(dp^2 + dq^2)/E^2$ maps to the unit
sphere, plane or
 pseudo-sphere, $|S(r)| \neq 0$ is needed for a sensible mapping, and so
$S > 0$ is a reasonable choice.  In the cases $\epsilon = 0$ or $-1$, $E$
necessarily goes to zero at certain $(p, q)$ values where the mapping is
badly behaved.  For a well behaved $r$ coordinate, we do need to specify
 \begin{equation}
   \infty > \frac{(R' - R E'/E)^2}{(\epsilon + f)} > 0,
 \end{equation}
 \begin{equation}
   \mbox{i.e.~~~~} (\epsilon + f) > 0 \mbox{~~except where~~}
   (R' - R E'/E)^2 = 0.   \label{epsf>0}
 \end{equation}
 In Lema\^{\i}tre-Tolman models \cite{Lem33, Tol34}
 ($E' = 0,~~ \epsilon = 1$), the equality $(1 + f) = 0 = (R')^2$ can occur
in closed models where the areal radius on a spatial section is at a
maximum, or in wormhole models where the areal radius is minimum, $R'(t,
r_m) = 0,~~ \forall~t$. These can only occur at constant $r$ and must hold
for all $(p,q)$ values.  We will consider maxima and minima again later.

 \item   The density must be positive, and the Kretschmann scalar must be
finite, which adds
 \begin{eqnarray}
    \mbox{either~~~~} M' - 3 M E' / E \geq 0 ~~~~~~~~ \mbox{and} ~~~~~~~~
      R' - R E' / E \geq 0  \label{MrErRrp} \\ \nonumber \\
   \mbox{or~~~~} M' - 3 M E' / E \leq 0 ~~~~~~~~ \mbox{and} ~~~~~~~~
      R' -R E' / E \leq 0.  \label{MrErRrm}
 \end{eqnarray}

     If $(R' - R E'/E)$ passes through $0$ anywhere other than a regular
extremum, we have a shell crossing, where an inner shell of matter passes
through an outer shell, and the density diverges and goes negative.  This
phenomenon is probably due to the spacetime coordinates being attached to
the shells of matter, and is not physically realistic.  Nevertheless, we
would like to avoid models in which such unphysical behaviour occurs, so
it is useful to find restrictions on the arbitrary functions that prevent
it.

 \item   The various arbitrary functions should have sufficient continuity
 --- $C^1$ and piecewise $C^3$
 --- except possibly at a spherical origin.

 \end{enumerate}

 \section{The Significance of $E$}

 \subsection{Properties of $E(r, p, q)$.}\label{PoErpq}

 Note that the Szekeres metric is covariant with the transformations $r =
g(\tilde{r})$, where $g$ is an arbitrary function.  Hence, if $R' < 0$ in the
neighbourhood of some value $r = r_0$, we can take $g = 1/\tilde{r}$ and obtain
$dR/d\tilde{r} > 0$.  Therefore, $R' > 0$ can always be assumed to hold in some
neighbourhood of any $r = r_0$.  However, if $R'$ changes sign somewhere, then
this is a coordinate-independent property.

 As seen from eq. (\ref{Edef}), with $\epsilon = +1$, $E$ must be always
nonzero.  Since the sign of $E$ is not defined by the metric, we
can assume that $E > 0$.

 Can $E'$ change sign?
 \begin{equation}\label{defE'}
 E' = \frac 12 S' \left\{- \left[(p - P)^2 + (q - Q)^2\right]/S^2 +
\epsilon \right\} - \frac 1S [(p - P)P' + (q - Q)Q'].
 \end{equation}
 The discriminant of this with respect to $(p - P)$ is
 \begin{equation}\label{discpE'}
 \Delta_p = \frac 1{S^2}\left[- \frac {S'^2}{S^2} (q - Q)^2 - 2
\frac {S'}{S}(q - Q)Q' + P'^2 + \epsilon S'^2\right].
 \end{equation}
 The discriminant of $\Delta_p$ with respect to $(q - Q)$ is
 \begin{equation}\label{discqE'}
 \Delta_q = 4 \frac {S'^2}{S^{6}} (P'^2 + Q'^2 + \epsilon S'^2).
 \end{equation}
 Since, with $\epsilon = +1$, this is never negative, the equation
$E' = 0$ will always have at least one solution (exceptional
situations), and in general two.  The two exceptional situations
are when $\Delta_q = 0$.  They are:
 \begin{enumerate}
 \item   $S' = 0$. Then $E' = 0$ has a family of solutions anyway, but
the solutions define a straight line in the $(p, q)$-plane. This will be dealt
with below (see after eq. (\ref{E'=0TanTheta})).
 \item   $S' = P' = Q' = 0$.  Then $E' \equiv 0$ at this particular value
of $r$, and we see from eq. (\ref{RhoDef}) that $\rho$ will be spherically
symmetric there. (In this case, the positions of the great circle from eq.
(\ref{E'0eq}) and of the poles from eq. (\ref{E'Eextreme}) are undetermined).
 \end{enumerate}

\begin{figure}
 \begin{center}
 \parbox{14cm}{
 \includegraphics[scale = 0.75]{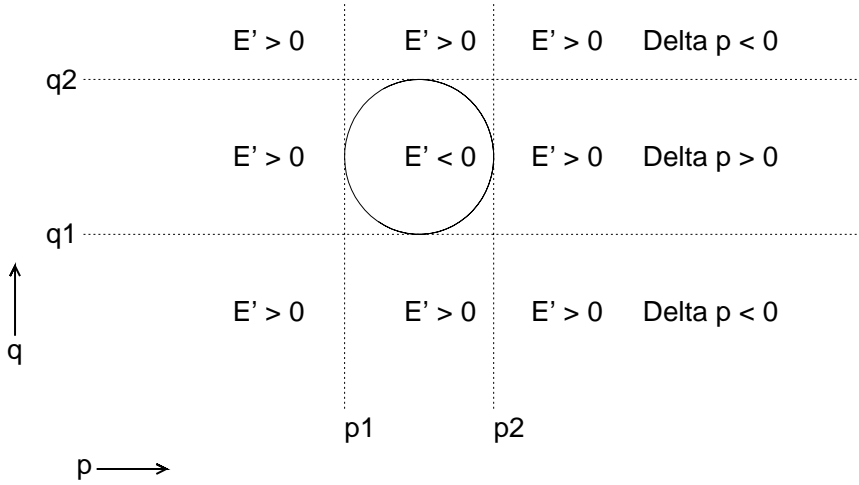}
 \caption {
 \label{E'plane}
 \footnotesize
 When $S' > 0$, $E' < 0$ inside the circle defined by (\ref{eqE'circ}) and
$E' > 0$ outside.  Note that the $(p, q)$ plane is a Riemann projection of a
sphere, and on the sphere "inside" and "outside" are topologically equivalent.
 }
 }
 \end{center}
 \end{figure}

 When $\Delta_q > 0$, $\Delta_p$ will change sign at the following
two values of $q$:
 \begin{equation}\label{solDelp}
 q_{1,2} = Q + \frac {S}{S'}\left(- Q' \pm \sqrt{P'^2 + Q'^2 + \epsilon
S'^2}\right).
 \end{equation}
 For every $q$ such that $q_1 < q < q_2$ there will be two values
of $p$ (and one value of $p$ when $q = q_1$ or $q = q_2$) such
that $E' = 0$.  Those values of $p$ are
 \begin{equation}\label{solE'0}
 p_{1,2} = P - P'\frac {S}{S'} \pm S\sqrt{- \left(\frac {q - Q}S +
\frac {Q'}{S'}\right)^2 + \frac {P'^2 + Q'^2}{S'^2} + \epsilon}.
 \end{equation}
 The regions where $E'$ is positive and negative depend on the sign of $S'$.  If
$S' > 0$, then $E' > 0$ for $p < p_1$ and for $p > p_2$, if $S' < 0$, then $E'
> 0$ for $p_1 < p < p_2$. $E' = 0$ for $p = p_1$ and $p = p_2$, but note that
$p_1$ and $p_2$ are members of a continuous family labelled by $q$.  All the
values of $p$ and $q$ from (\ref{solDelp}) -- (\ref{solE'0}) lie on the circle
 \begin{equation}\label{eqE'circ}
 \left[p - \left(P - P' \frac {S}{S'}\right)\right]^2 +\left[q -
\left(Q - Q' \frac {S}{S'}\right)\right]^2 = S^2\left(\frac{P'^2 +
Q'^2}{S'^2} + \epsilon\right).
 \end{equation}
 The center of this circle is in the point
 \begin{equation}\label{cenE'circ}
 (p, q) = \left(P - P'\frac{S}{S'} , Q - Q'\frac{S}{S'}\right),
 \end{equation}
 and the radius of this circle is
 \begin{equation}\label{radE'circ}
 {L_{E'=0}} = S\sqrt{\frac{P'^2 + Q'^2}{S'^2} + \epsilon}.
 \end{equation}
 The situation on the $(p,q)$-plane when $S' > 0$ is shown in Fig. \ref{E'plane}.

 \bigskip

 \subsection{Properties of $E(r, \theta, \phi)$.}

     We consider the variation of $E(r, p, q)$ around the spheres of
constant $t$ and $r$.

     Setting $\epsilon = +1$ and applying the transformation
(\ref{Riemprojp}) to (\ref{Edef}) and to its derivative gives
 \begin{eqnarray}
   E   & = & \frac{S}{1 - \cos \theta},   \label{E_theta}
   \\ \nonumber \\
   E'  & = & - \frac{S' \cos \theta +
                        \sin \theta (P' \cos \phi + Q' \sin \phi)}
                    {1 - \cos \theta},   \label{E'_thetaphi}
   \\ \nonumber \\
   E'' & = & - \frac{S'' \cos \theta + \sin \theta (P'' \cos \phi
                + Q'' \sin \phi)}
                    {(1 - \cos \theta)}
   \nonumber \\ \nonumber \\
       && + 2 \left( \frac{S'}{S} \right) \left( \frac{S' \cos \theta
             + \sin \theta (P' \cos \phi + Q' \sin \phi)}
             {(1 - \cos \theta)} \right)
   \nonumber \\ \nonumber \\
       && + \frac{(S')^2 + (P')^2 + (Q')^2}{S}.
 \end{eqnarray}

 The locus $E' = 0$ is
 \begin{equation}
   S' \cos \theta + P' \sin \theta \cos \phi + Q' \sin \theta \sin \phi
      = 0.   \label{E'0eq}
 \end{equation}
 Writing $z = \cos \theta$, $y = \sin \theta \cos \phi$, $x = \sin \theta
\sin \phi$, clearly puts $(x, y, z)$ on a unit sphere through $(0, 0, 0)$,
and (\ref{E'0eq}) becomes $S' z + P' x + Q' y = 0$ which is the equation
of an arbitrary plane through $(0, 0, 0)$.  Such planes all intersect the
unit sphere along great circles, therefore $E' = 0$ is a great circle,
with locus
 \begin{equation}   \label{E'=0TanTheta}
   \tan \theta = \frac{- S'}{P' \cos \phi + Q' \sin \phi}.
 \end{equation}
 The plane has unit normal $(P', Q', S')/\sqrt{(P')^2 + (Q')^2 +
(S')^2}\;$.

Now it is easy to understand the meaning of the special case $S' = 0$ mentioned
after eq. (\ref{discqE'}). As seen from (\ref{E'=0TanTheta}), with $S' = 0$ we have
$\theta = 0$, which means that the great circle defined by $E' = 0$ passes
through the pole of Riemann projection. In this case, the image of the circle
$E' = 0$ on the $(p, q)$ plane is a straight line passing through $(p, q) = (P,
Q)$, as indeed follows from (\ref{defE'}). The sign of $E'$ is different on
each side of the straight line. Compare also with Figs. \ref{proj1fig} and
\ref{proj2fig}.

     From (\ref{E'_thetaphi}) and (\ref{E_theta}) we find
 \begin{equation}
   \frac{E'}{E} = - \frac{S' \cos \theta +
                    \sin \theta (P' \cos \phi + Q' \sin \phi)}{S}
   \label{ErEthetaphi}
 \end{equation}
 thus
 \begin{equation}   \label{E'/Econst}
   \frac{E'}{E} = \mbox{constant} ~~~~\Rightarrow~~~~
   S' z + P' x + Q' y = S \times \mbox{constant}
 \end{equation}
 which is a plane parallel to the $E' = 0$ plane, implying that all loci
$E' / E =$~constant are small circles parallel to the $E' = 0$ great
circle.  This will be seen to apply to shell crossings and apparent
horizons.

 The location of the extrema of $E'/E$ are found as follows
 \begin{eqnarray}
   \frac{\partial (E'/E)}{\partial \phi} & = &
      \frac{\sin \theta (P' \sin \phi - Q' \cos \phi)}{S} = 0
      ~~~~\Rightarrow~~~~   \\ \nonumber \\
   \tan \phi_e & = & \frac{Q'}{P'}
      ~~~~\Rightarrow~~~~ \cos \phi_e = \epsilon_1 \frac{P'}{\sqrt{(P')^2
      + (Q')^2}\;}, \qquad \epsilon_1 = \pm 1,   \label{TanPhiX}   \\ \nonumber \\
   \frac{\partial (E'/E)}{\partial \theta} & = &
      \frac{S' \sin \theta - P' \cos \theta \cos \phi -
      Q' \cos \theta \sin \phi }{S} = 0
      ~~~~\Rightarrow~~~~   \\ \nonumber \\
   \tan \theta_e & = & \frac{P' \cos \phi_e + Q' \sin \phi_e}{S'}
      = \epsilon_1 \frac{\sqrt{(P')^2 + (Q')^2}\;}{S'}
      ~~~~\Rightarrow~~~~   \label{TanThetaX}   \\ \nonumber \\
   \cos \theta_e & = & \epsilon_2 \frac{S'}{\sqrt{(S')^2 + (P')^2 + (Q')^2}\;},
   \qquad \epsilon_2 = \pm 1.
 \end{eqnarray}
 The extreme value is then
 \begin{equation}
   \left( \frac{E'}{E} \right)_{\rm extreme} =
      - \epsilon_2 \frac{\sqrt{(S')^2 + (P')^2 + (Q')^2}}{S}.
      \label{E'Eextreme}
 \end{equation}
 Since $(\sin \theta_e \cos \phi_e, \sin \theta_e \sin \phi_e, \cos
\theta_e) = \epsilon_2 (P', Q', S')/\sqrt{(P')^2 + (Q')^2 + (S')^2}\;$, eq.
(\ref{E'0eq}) shows that the extreme values of $E'/E$ are poles to the great
circles of $E' = 0$.  The latter can now be written in parametric form as
 \begin{equation}
   \cos \theta = - \cos \psi \sin \theta_e ~~,~~~~~~
 \end{equation}
 \begin{equation}
   \tan \phi = \frac{\cos \theta_e \tan \phi_e + \tan \psi}
      {\cos \theta_e - \tan \phi_e \tan \psi}
 \end{equation}

     Clearly $E'/E$ has a dipole variation around each constant $r$
sphere, changing sign when we go over to the anitipodal point: $(\theta, \phi)
\to (\pi - \theta, \phi + \pi)$. Writing
 \begin{equation}
   \left( R' - R \frac{E'}{E} \right) =
   \left( R' + R \frac{S' \cos \theta +
      \sin \theta (P' \cos \phi + Q' \sin \phi)}{S} \right)
 \end{equation}
 we see that $R E'/E$ is the correction to the radial separation $R'$ of
neighbouring constant $r$ shells, due to their not being concentric.  In
partuicular $R S' / S$ is the forward $(\theta = 0)$ displacement, and $R
P' / S$ \& $R Q' / S$ are the two sideways displacements $(\theta =
\pi/2,~ \phi = 0)$ \& $(\theta = \pi/2,~ \phi = \pi/2)$.  The shortest
`radial' distance is where $E'/E$ is maximum.

It will be shown in section \ref{NoShCr} that, where $R' > 0$, $E'/E \leq
M'/(3M)$ and $E'/E \leq R'/R$ are required to avoid shell crossings, and also
in eq (\ref{R'/R > M'/3M}) that $R'/R > M'/3M$. These inequalities, together
with $M'
> 0$, imply that the density given by (\ref{RhoDef}), as a function of $x =
E'/E$:

\begin{equation}\label{rhoofx}
\rho = \frac {2M'}{R^2R'} \frac {1 - 3Mx/M'}{1 - Rx/R'}
\end{equation}
has a negative derivative by $x$:

\begin{equation}\label{derxrho}
\rho,_x = \frac {R/R' - 3M/M'}{(1 - Rx/R')^2}\cdot\frac{2M'}{R^2R'} < 0,
\end{equation}
and so the density is minimum where $E'/E$ is maximum.

The density, eq. (\ref{RhoDef}), can be decomposed into a spherical part and a
dipole-like part, as noted by Szekeres \cite{Sze75b} and de Souza \cite{deS85}
(see also p. 30 in Ref.\cite{Kra97}). Rewriting de Souza's result into our
notation of eqs. (\ref{dsS}) and (\ref{Edeforig}), we obtain:

\begin{equation}\label{rhosplit}
\rho = \rho_s + \Delta\rho,
\end{equation}
where $\rho_s$ is the spherical part:

\begin{equation}\label{rhosphepa}
\rho_s = \frac {2M'(A + C) - 6M(A' + C')}{R^2[R'(A + C) - R(A' + C')]},
\end{equation}
and $\Delta\rho$ is the dipole-like part:

\begin{equation}\label{diprho}
\Delta\rho = \frac {A' + C' - (A + C)E'/E}{R' - RE'/E} \cdot \frac {6MR' -
2M'R}{R^2[R'(A + C) - R(A' + C')]}.
\end{equation}
The dipole-like part changes sign on the surface where $E'/E = (A' + C')/(A +
C)$, but lacks the antisymmetry property: $\Delta\rho(-E'/E) \neq -
\Delta\rho(E'/E)$. It can be verified (see Appendix \ref{massdipole}) that the
$\Delta\rho = 0$ hypersurface does intersect every $(t = {\rm const}, r = {\rm
const})$ sphere along a circle, unless $P' = Q' = S' = 0$ ($= A' = C'$), in
which case the dipole component of density is simply zero. The surface
$\Delta\rho = 0$ in a $t =$ const space is comoving, i.e. its definition does
not depend on $t$. Also, its intersection with any sphere of constant $r$,
$E'/E = (A' + C')/(A + C) =$ const, is a circle parallel to the great circle
$E' = 0$, as noted after eq. (\ref{E'/Econst}). It will coincide with the $E' = 0$
circle in those points where $A' + C' = 0$ (if they exist). The dipole-like
component will be antisymmetric with respect to $E'/E$ only at such values of
$r$, where $(A + C)R = 0 = (A' + C')R'$, but such values may exist only at the
center, $R = 0$, because $A + C = 0$ contradicts eq. (\ref{Econd}).

 In the maximum ($\epsilon_2 = -1$) and minimum ($\epsilon_2 = 1$) directions,
 \begin{equation}
   E''_{\rm max/min} = 
 $$ $$
 \frac{ S (S'' S' + P'' P' + Q'' Q')
   - [(S')^2 + (P')^2 + (Q')^2](2S' + \epsilon_2 \sqrt{(S')^2 + (P')^2 +
   (Q')^2}\;)   }
   { S (S' - \epsilon_2 \sqrt{(S')^2 + (P')^2 + (Q')^2}\;) },
 \end{equation}
 while around the $E' = 0$ circle
 \begin{equation}
   E'' = \frac{(S')^2 + (P')^2 + (Q')^2}{S}
   - \frac{S'' (P' \cos \phi + Q' \sin \phi)
   - S' (P'' \cos \phi + Q'' \sin \phi)}
   {\sqrt{(P' \cos \phi + Q' \sin \phi)^2 + (S')^2}\;
   - (P' \cos \phi + Q' \sin \phi)}.
 \end{equation}

 \section{Regular Origins}
 \label{RegOrig}

     When $\epsilon = +1$, $R = 0$ occurs at an origin of spherical
coordinates, e.g. $R(t, 0) = 0,~~\forall~t$, where the
 2-spheres have no size.  Similarly, $\dot{R}(t, 0) = 0 = \ddot{R}(t, 0)$,
etc $\forall~t$.  There will be a second origin, at $r = r_O$ say, in any
closed, regular, $f < 0$ model.  Thus, by (\ref{ellevRS}) and (\ref{hypevRS})
and their combinations with (\ref{RdotSq}) \& (\ref{Rddot}), for each constant
$\eta$
 \begin{equation}
   \lim_{r \rightarrow 0} \frac{M}{f} = 0
   ~~,~~~~~~
   \lim_{r \rightarrow 0} f = 0
   ~~,~~~~~~
   \lim_{r \rightarrow 0} \frac{f^2}{M} = 0.
 \end{equation}

 The type of time evolution at the origin must be the same as its
neighbourhood, i.e., along a constant $t$ slice away from the bang or crunch,
by (\ref{ellevtS}) and (\ref{hypevtS}),
 \begin{equation}
   0 < \lim_{r \rightarrow 0} \frac{|f|^{3/2} (t - a)}{M} < \infty.
 \end{equation}
 Clearly, we need $M \rightarrow 0$, $f \rightarrow 0$ and
 \begin{equation}\label{Limf32M}
   0 < \lim_{r \rightarrow 0} \frac{|f|^{3/2}}{M} < \infty.
 \end{equation}
 Using l'H\^{o}pital's rule, this gives
 \begin{equation}
   \lim_{r \rightarrow 0} \frac{3 M f'}{2 M' f} = 1.
   \label{lHopLimf32M}
 \end{equation}

 The density and Kretschmann scalar must be well behaved.  We don't
consider a vacuum region of finite size at the origin, as that is just
Minkowski space, $M = 0$, and we don't consider the obscure case of a
single vacuum point at the origin.  Because $\rho$ \& $\overline{\rho}$
in (\ref{KretschDef}) evolve differently, we also need
 \begin{equation}
   0 < \lim_{r \rightarrow 0} \frac{6 M}{R^3}
      = \lim_{r \rightarrow 0} \frac{2 M'}{R^2 R'}
      < \infty
   ~~~~\Rightarrow~~~~
   \lim_{r \rightarrow 0} \frac{3 R' M}{R M'} = 1
   \label{rhorhobarlim1}
 \end{equation}
 and
 \begin{equation}
   0 < \lim_{r \rightarrow 0} \frac{2 (M' - 3 M E'/E)}{R^2 (R' - R E'/E)}
     = \lim_{r \rightarrow 0}
       \frac{2 M'}{R^2 R'} \frac{(1 - 3 M E'/M'E)}{(1 - R E'/R'E)}
     < \infty,
 \end{equation}
 but in fact the latter is ensured by the former, and the anisotropic
effect of $E$ vanishes at the origin.  However, since $E'/E$ is restricted
by the conditions for no shell crossings, it would be odd if $\lim_{r
\rightarrow 0} M E'/M' E$ or $\lim_{r \rightarrow 0} R E'/R' E$ were
divergent.  Since
 \begin{eqnarray}
   R^2 \frac{R'}{M'} & = &
   - \frac{M^2}{(-f)^3}
     \left( 1 - \frac{3 M f'}{2 M' f} \right)
     \sin \eta (\eta - \sin \eta) (1 - \cos \eta)
     \nonumber \\ \nonumber \\
   && + \frac{M^2}{(-f)^3}
     \left( 1 - \frac{M f'}{M' f} \right)
     (1 - \cos \eta)^3
     \nonumber \\ \nonumber \\
   && - \frac{M^2 a'}{(-f)^{3/2} M'}
     \sin \eta (1 - \cos \eta),
 \end{eqnarray}
eqs. (\ref{Limf32M}) and (\ref{lHopLimf32M}) above make the first term zero and
the second non-zero at an origin for all $0 < \eta < 2 \pi$, so we only need
 \begin{equation}
   \lim_{r \rightarrow 0} \frac{M a'}{M'} < \infty.
 \end{equation}

 Lastly, the metric must be well behaved, so $E$ should have no unusual
behaviour, such as $S = 0$, that would compromise a valid mapping of
$(dp^2 + dq^2)/E^2$ to the unit sphere.  Also, to ensure the rate of
change of proper radius with respect to areal radius is that of an origin,
$g_{rr}/(R')^2$ should be finite
 \begin{eqnarray}
   && 0 < \lim_{r \rightarrow 0}
          \frac{(R' - R E'/E)^2}{(1 + f) (R')^2}
        < \infty \\ \nonumber \\
   \Rightarrow~~
   && 0 < \lim_{r \rightarrow 0}
          \left( 1 - \frac{3 M E'}{M' E} \right)^2
        < \infty
   \Rightarrow~~
      - \infty \leq \lim_{r \rightarrow 0}
             \left| \frac{M E'}{M' E} \right|
        <    \infty \\ \nonumber \\
   && {\rm and} \qquad \lim_{r \rightarrow 0}
             \left| \frac{M E'}{M' E} \right| \neq \frac 13,
 \end{eqnarray}
 where the last of (\ref{rhorhobarlim1}) has been used.  This should hold
for all $(p, q)$, i.e. all $(\theta, \phi)$.  Thus (\ref{ErEthetaphi})
gives
 \begin{equation}
      - \infty \leq \lim_{r \rightarrow 0}
             \left| \frac{M S'}{M' S} \right|
        <    \infty
      ~~,~~~~~~
      - \infty \leq \lim_{r \rightarrow 0}
             \left| \frac{M P'}{M' S} \right|
        <    \infty
      ~~,~~~~~~
      - \infty \leq \lim_{r \rightarrow 0}
             \left| \frac{M Q'}{M' S} \right|
        <    \infty,
 \end{equation}
all three limits being different from $1/3$.

 All of the above suggests that, near an origin,
 \begin{equation}
   M \sim R^3 ~~,~~~~
   f \sim R^2 ~~,~~~~
   S \sim R^n ~,~~~~
   P \sim R^n ~,~~~~
   Q \sim R^n ~,~~~~ n \geq 0.
 \end{equation}
 The condition $E'/E \leq M'/3M$ that will be obtained in the next section
implies that near an origin
 \begin{equation}
   n \leq 1
 \end{equation}

 \section{Shell Crossings}

 \subsection{Occurrence and Position of Shell Crossings in a
             Surface of Constant $t$ and $r$.}\label{OaPoSCiaSoCtar}

 A shell crossing, if it exists, is the locus of zeros of the function $R' -
RE'/E$.  Suppose that $R' - RE'/E = 0$ holds for all $r$ at some $t = t_{o}$.
This leads to $S' = P' = Q' = R' = 0$.  Since $P$, $Q$ and $S$ depend only on
$r$, this means they are constant throughout the spacetime.  As seen from
(\ref{dsS}) and (\ref{RhoDef}), the Szekeres metric reduces then to the LT
metric, and so this case need not be considered.

 Suppose that $R' - RE'/E = 0$ holds for all $t$ at some $r = r_{0}$.  This is
an algebraic equation in $p$ and $q$ whose coefficients depend on $t$ and $r$.
Taking the coefficients of different powers of $p$ and $q$ we find $P' = Q' =
S' = R' = 0$, but this time these functions vanish only at $r = r_0$, while
$R'(t,r_0)$ will vanish for all $t$.  This will either be a singularity (when
$M'(r_0) \neq 0$) or a neck (when $M'(r_0) = 0$), familiar from the studies of
the LT model, see Refs. \cite{Hel87} and \cite{HelLak85}.  Hence, $R' -
RE'/E \neq 0$ except at a shell crossing or at special locations.

 Now $R' > 0$ and $R' - RE'/E < 0$ cannot hold for all $p$ and $q$.
This would lead to $E' > ER'/R > 0$, and we know that $E'$ cannot
be positive at all $p$ and $q$.  Hence, with $R' > 0$, there must be a
region in which $R' - RE'/E > 0$.  By a similar argument, $R' < 0$ and $R'
- RE'/E > 0$ cannot hold for all $p$ and $q$, so with $R' < 0$, there must
be a region in which $R' - RE'/E < 0$.

 Assuming $R' > 0$, can $R' - RE'/E$ be positive for all $p$ and
$q$?  Writing
 \begin{equation}\label{defE-E'}
 ER'/R - E' = \frac 1{2S}\left(\frac{S'}S + \frac {R'}R\right)
\left[(p - P)^2 + (q - Q)^2\right] - \frac 12 \epsilon
S\left(\frac {S'}S - \frac {R'}R \right)
 $$ $$
 + \frac 1S [(p - P)P' + (q - Q)Q'],
 \end{equation}
 the discriminants of this with respect to $(p - P)$ and $(q - Q)$ are
 \begin{eqnarray}
 \label{discpSC}
 \Delta_p & = & \frac {P'^2}{S^2} - \frac 1{S^2} \left(\frac{S'}S +
\frac {R'}R\right)\left[\left(\frac{S'}S + \frac {R'}R\right)(q -
Q)^2 + (q - Q)Q' - \frac 12 \epsilon \left(\frac{S'}S - \frac
{R'}R\right)\right] \\
 \label{discqSC}
 \Delta_q & = & 4\frac 1{S^2}\left(\frac{S'}S + \frac {R'}R \right)^2
\left[\frac{P'^2 + Q'^2 + \epsilon S'^2}{S^2} - \epsilon \frac
{R'^2}{R^2} \right].
 \end{eqnarray}
 Thus $ER'/R - E'$ will have the same sign for all $p$ and $q$ when
$\Delta_q < 0$ (because then also $\Delta_p < 0$ for all $q$).
Hence, $ER'/R - E'$ has the same sign for all $p$ and $q$ (i.e. there are
no shell crossings) if and only if
 \begin{equation}\label{noshcr}
 \frac {R'^2}{R^2} > \epsilon \frac {P'^2 + Q'^2 + \epsilon
S'^2}{S^2} := \Phi^2(r).
 \end{equation}
Note that when $\epsilon = 0$, this can fail only at those points where $R' =
0$.

 If $R'^2/R^2 = \Phi^2$, then $\Delta_q = 0$, and so $\Delta_p = 0$
at just one value of $q = q_{SS}$.  At this value of $q$, $ER'/R -
E' = 0$ at one value of $p = p_{SS}$.  In this case, the
shell crossing is a single point in the constant $(t, r)$-surface,
i.e. a curve in a space of constant $t$ and a 2-surface in
spacetime.

 If $R'^2/R^2 < \Phi^2$, then the locus of $ER'/R - E' = 0$ is in
general a circle (a straight line in the special case $S'/S = -
R'/R$) in the $(p, q)$ plane.  The straight line is just a
projection onto the $(p, q)$ plane of a circle on the sphere of
constant $t$ and $r$, and so is not really any special case.

 When $\Delta_q > 0$ ($R'^2/R^2 < \Phi^2$), the two limiting values of $q$ at
which $\Delta_p$ changes sign are
 \begin{equation}\label{defqSC}
 q_{1,2} = \frac {- Q' \pm \sqrt{P'^2 + Q'^2 + \epsilon \left(S'^2
- S^2R'^2/R^2\right)}} {S'/S + R'/R},
 \end{equation}
 and then for every $q$ such that $q_1 < q < q_2$, there are two
values of $p$ (only one if $q = q_1$ or $q = q_2$) such that
$ER'/R - E' = 0$.  These are
 \begin{equation}\label{defpSC}
 p_{1,2} = \frac {- P' \pm \sqrt{- \left[\left(\frac{S'}S + \frac
{R'}R\right)(q - Q) - Q'\right]^2 + P'^2 + Q'^2 + \epsilon
\left(S'^2 - S^2R'^2/R^2\right)}} {S'/S + R'/R}.
 \end{equation}
 The values of $p$ and $q$ from (\ref{defqSC}) and (\ref{defpSC})
lie on the circle with the center at
 \begin{equation}\label{cenSC}
 \left(p_{SC}, q_{SC}\right) = \left(P - \frac {P'}{S'/S + R'/R}, Q
- \frac {Q'}{S'/S + R'/R}\right),
 \end{equation}
 and with the radius $L_{SC}$ given by
 \begin{equation}\label{radSC}
 {L_{SC}}^2 = \frac {P'^2 + Q'^2 + \epsilon\left(S'^2 -
S^2R'^2/R^2\right)}{(S'/S + R'/R)^2}.
 \end{equation}
 This is in general a different circle than the one defined by $E' = 0$.  As seen
from (\ref{defE-E'}), the shell crossing set intersects with the surface of
constant $t$ and $r$ along the line $E'/E = R'/R =$ const.  As noted after eq.
(\ref{E'/Econst}), this line is a circle that lies in a plane parallel to the $E' =
0$ great circle. It follows immediately that the $E' = 0$ and the SC circles
cannot intersect unless they coincide.

 \subsection{Conditions for No Shell Crossings}\label{CfNSC}
 \label{NoShCr}

     Szekeres \cite{Sze75b}
 obtained a number of regularity conditions for the $\epsilon = +1$
metric, namely:~~ (1) On any constant time slice, $R(t=const,r)$ is monotonic
in $r$, which allows a transformation to make $R = r$ and $R' = 1$ on that
slice.~~ (2) At an origin, $A$, $B_1$, $B_2$ \& $C$ should be $C^1$, $f = 0$,
and $M \sim R^2$ but we are not sure why he required $A' = 0 = B_1' = B_2' =
C'$ there.~~ (3) To keep the density
 non-singular, $0 \leq ((S')^2 + (P')^2 + (Q')^2)/S^2 < min((R'/R)^2,
(M'/3M)^2)$, which is a no shell crossing condition.  We shall improve on
the latter below.

     For positive density, (\ref{RhoDef}) shows that $(M' - 3 M E'/E)$ \&
$(R' - R E'/E)$ must have the same sign.  We now consider the case where
both are positive.  Where $(M' - 3 M E'/E) \leq 0$ and $(R' - R E'/E) < 0$
we reverse the inequalities in all the following.

 In the case of both $(M' - 3 M E'/E)$ \& $(R' - R E'/E)$ being zero, this
can hold for a particular $(p, q)$ value if $M' / 3 M = R' / R$, but the
latter cannot hold for all time.  This case can only hold for all $(p, q)$
if $M' = 0$, $E' = 0$, $R' = 0$, which requires all of $M'$, $f'$, $a'$,
$S'$, $P'$, $Q'$ to be zero at some $r$ value.

     We consider the inequality $(M' - 3 M E'/E) \geq 0$ and we argue that
it must hold even for the extreme value of $E'/E$, (\ref{E'Eextreme}), for
which we obtain
 \begin{equation}
   \frac{M'}{3M} \geq \left. \frac{E'}{E} \right|_{max}
    = \frac{\sqrt{(S')^2 + (P')^2 + (Q')^2}}{S}
    ~~~~\forall~r.  \label{MrCond}
 \end{equation}
 It is obvious that this is sufficient, and also that
 \begin{equation}
   M' \geq 0 ~~~~\forall~r.
 \end{equation}

We will now consider $(R' - R E' / E) > 0$ for all 3 types of evolution.

 \subsubsection{Hyperbolic evolution, $f > 0$}

For hyperbolic models, we can write:
 \begin{equation}
   \frac{R'}{R} = \frac{M'}{M}(1 - \phi_4) +
   \frac{f'}{f} \left( \frac{3}{2}\phi_4 - 1 \right) -
   \frac{f^{3/2} a'}{M} \phi_5,
   \label{R'RH}
 \end{equation}
 where
 \begin{equation}
   \phi_4 = \frac{\sinh \eta (\sinh \eta - \eta)}{(\cosh \eta - 1)^2}
~~,~~~~~~~~ \phi_5 = \frac{\sinh \eta}{(\cosh \eta - 1)^2}.
 \end{equation}
 At early times,
 \begin{eqnarray}
   \eta & \rightarrow & 0,
   \\ \nonumber \\
   R & \rightarrow & \frac{M}{f} \frac{\eta^2}{2} + O(\eta^4)
\rightarrow 0,
   \\ \nonumber \\
   \phi_5 & \rightarrow & \frac{4}{\eta^3} + O(\eta) \rightarrow + \infty,
   \\ \nonumber \\
   \phi_4 & \rightarrow & \frac{2}{3} + O(\eta^2) \rightarrow \frac{2}{3},
 \end{eqnarray}
 we find $\phi_5$ dominates and
 \begin{equation}
   \frac{R'}{R} ~~ \rightarrow ~~ - \frac{f^{3/2} a'}{M} \phi_5,
 \end{equation}
 so that $(R' - R E' / E) > 0$ gives
 \begin{equation}
   a' < 0 ~~~~\forall~r.  \label{arCondH}
 \end{equation}
 Similarly, at late times,
 \begin{equation}
   \eta \rightarrow \infty ~~,~~~~~~
   R \rightarrow \infty ~~,~~~~~~
   \phi_5 \rightarrow 0 ~~,~~~~~~
   \phi_4 \rightarrow 1,
 \end{equation}
 we find $\phi_5$ vanishes and
 \begin{equation}
   \frac{R'}{R} ~~ \rightarrow ~~ \frac{1}{2}\frac{f'}{f},
 \end{equation}
 so that
 \begin{equation}
   \left( \frac{R'}{R} - \frac{E'}{E} \right) > 0 ~~~~ \Rightarrow ~~~~
\frac{f'}{2f} - \frac{E'}{E} > 0.\label{R'/R>E'/E}
 \end{equation}
 Following the above analysis of  $(M' - 3 M E'/E) \geq 0$ we obtain
 \begin{equation}
   \frac{f'}{2f} > \frac{\sqrt{(S')^2 + (P')^2 + (Q')^2}}{S}
   ~~~~\forall~r,   \label{frErCondH}
 \end{equation}
 which obviously implies
 \begin{equation}
   f' > 0 ~~~~\forall~r.   \label{frCondH}
 \end{equation}
 Again, since we already have $M' \geq 0$, it is clear that this is
sufficient, and that
 \begin{equation}\label{posR'hyp}
   R' > 0.
 \end{equation}

 \subsubsection{Parabolic evolution, $f = 0$}

 The easiest way to obtain the conditions for this case, $f = 0$, $f' \neq
0$, is to put $\tilde{\eta} = \eta/\sqrt{f}\; > 0$ in the hyperbolic case,
and take the limit $f \rightarrow 0$, $\eta \rightarrow 0$.  All terms
involving $f'/f$ cancel and we retain exactly the same conditions, viz
(\ref{arCondH}) \& (\ref{frCondH}) (and of course (\ref{MrCond})).
Naturally, (\ref{frErCondH}) ceases to impose any limit.

 \subsubsection{Elliptic evolution, $f < 0$}

For elliptic models, we can write:
 \begin{equation}
   \frac{R'}{R} = \frac{M'}{M}(1 - \phi_1)
   + \frac{f'}{f} \left( \frac{3}{2}\phi_1 - 1 \right)
   - \frac{(-f)^{3/2} a'}{M} \phi_2,
   \label{R'RE}
 \end{equation}
 where
 \begin{equation}
   \phi_1 = \frac{\sin \eta (\eta - \sin \eta)}{(1 - \cos \eta)^2}
   ~~,~~~~~~~~ \phi_2 = \frac{\sin \eta}{(1 - \cos \eta)^2}.
   \label{phi1phi2}
 \end{equation}
 At early times,
 \begin{eqnarray}
   \eta & \rightarrow & 0,
   \\ \nonumber \\
   R & \rightarrow & \frac{M}{(-f)} \frac{\eta^2}{2} + O(\eta^4)
\rightarrow 0,
   \\ \nonumber \\
   \phi_2 & \rightarrow & \frac{4}{\eta^3} + O(\eta) \rightarrow + \infty,
   \\ \nonumber \\
   \phi_1 & \rightarrow & \frac{2}{3} + O(\eta^2) \rightarrow \frac{2}{3},
 \end{eqnarray}
 we find $\phi_2$ dominates and
 \begin{equation}
   \frac{R'}{R} ~~ \rightarrow ~~ - \frac{f^{3/2} a'}{M} \phi_2,
 \end{equation}
 so that $R (R' / R - E' / E) > 0$ gives
 \begin{equation}
   a' < 0 ~~~~\forall~r.  \label{arCondE}
 \end{equation}
 Similarly, at late times,
 \begin{eqnarray}
   \eta & \rightarrow & 2 \pi,
   \\ \nonumber \\
   R & \rightarrow & \frac{M}{(-f)} \frac{(2 \pi - \eta)^2}{2} + O((2 \pi
- \eta)^4) \rightarrow 0,
   \\ \nonumber \\
   \phi_2 & \rightarrow & - 4 / (2 \pi - \eta)^3 + O((2 \pi - \eta))
      \rightarrow - \infty, \\
   \phi_1 & \rightarrow & - 8 \pi / (2 \pi - \eta)^3 + 2/3  + O((2 \pi -
      \eta)) \rightarrow - \infty,
 \end{eqnarray}
 we find
 \begin{equation}
   \frac{R'}{R} ~~ \rightarrow ~~ \frac{M'}{M}
   \left(\frac{8 \pi}{(2 \pi - \eta)^3} \right)
   - \frac{f'}{f} \left( \frac{12 \pi}{(2 \pi - \eta)^3} \right)
   + \frac{(-f)^{3/2} a'}{M} \left( \frac{4}{(2 \pi - \eta)^3} \right),
 \end{equation}
 so that $R^{3/2} (R' / R - E' / E) > 0$ now gives
 \begin{equation}
   \frac{2 \pi M}{(-f)^{3/2}}
   \left( \frac{M'}{M} - \frac{3 f'}{2 f} \right) + a' > 0 ~~~~\forall~r,
   \label{crunchCondE}
 \end{equation}
which is the condition that the crunch time must increase with $r$.  Since we
already have $M' \geq 0$, it may be easily verified that these conditions are
sufficient to keep
 \begin{equation}
   R' > 0
 \end{equation}
 for all $\eta$.

 We now show the above also ensure $R (R' / R - E' / E) > 0$ always.
Defining the crunch time $b(r)$ with
 \begin{equation}
   b = a + \frac{2 \pi M}{(-f)^{3/2}}
   \label{Def_b(r)}
 \end{equation}
 we can re-write (\ref{R'RE}) as
 \begin{equation}
   \frac{R'}{R} = \frac{M'}{3M}
   + \frac{b'}{(b - a)} \left( \frac{2}{3} - \phi_1 \right)
   + \frac{(-a')}{(b - a)} \left( \frac{2}{3} - \phi_1 + 2 \pi \phi_2 \right).
 \end{equation}
 The derivative of $(2/3 - \phi_1)$ is $(2 \eta - 3 \sin \eta + \eta \cos
\eta)/(1 - \cos \eta)^2$, and the third derivative of the numerator of the
latter is $\eta \sin \eta$.  It follows that $(2/3 - \phi_1) \geq 0$ and
declines monotonically from $+\infty$ to $0$ as $\eta$ goes from $2 \pi$
to $0$.  Since $(2/3 - \phi_1 + 2 \pi \phi_2)$ is the mirror image in
$\eta = \pi$ of $(2/3 -\phi_1)$, we have that
 \begin{equation}
   \frac{R'}{R} > \frac{M'}{3 M},
   \label{R'/R > M'/3M}
 \end{equation}
 so that (\ref{MrCond}) guarantees that for each given $r$, the maximum of
$E'/E$ as $(p,q)$ are varied is no more than the minimum of $R'/R$ as
$\eta$ varies.

 Note that although (\ref{crunchCondE}) implies
 \begin{equation}
   \frac{f'}{2 f} < \frac{M'}{3 M},
 \end{equation}
 a condition such as (\ref{frErCondH}) is not needed in this case.  As an
indication of the approximate magnitude of $R'/R|_{\rm min}$, at the moment of
maximum expansion along any given worldline,
 \begin{equation}
   \frac{R'}{R} = \frac{M'}{M} - \frac{f'}{f},
 \end{equation}
 so it would be possible to have $E'/E|_{\rm max}$ close to $R'/R|_{\rm
min}$ around the time of maximum expansion.

 \section{Regular Maxima \& Minima}\label{Regmaxmin}

     Certain topologies necessarily have extrema in $R$.  For example,
closed spatial sections have a maximum areal radius, and wormholes have a
minimum areal radius, i.e. $R'(t, r_m) = 0, ~~\forall~t$.

     Suppose $(\epsilon + f) = 0$ at some $r = r_m$.  By (\ref{epsf>0}) we
must have
 \begin{equation}
   f'(r_m) = 0
 \end{equation}
 (unless $f'$ is discontinuous there, which we won't consider).  We need
$(R' - R E' / E) = 0$ to keep $g_{rr}$ finite, and hence $(M' - 3 M E' /
E) = 0$ to keep $\rho$ finite, both holding $\forall~(t,p,q)$ at that
$r_m$.  More specifically,
 along any given spatial slice away from the bang or crunch,
 we want
 \begin{eqnarray}
   \frac{(R' - R  E' / E)}{\sqrt{\epsilon + f}\;} \rightarrow L
   ~~,~~~~~~~~ 0 < L < \infty,   \label{RegMaxLimR}
   \\ \nonumber \\
   \frac{(M' - 3 M E' / E)}{(R' - R  E' / E)} \rightarrow N
   ~~,~~~~~~~~ 0 \leq N < \infty.   \label{RegMaxLimM}
 \end{eqnarray}

 As noted above, we require
 \begin{equation}
   M' = f' = a' = S' = P' = Q' = 0   \label{RegMaxCond2}
 \end{equation}
 to ensure
 \begin{equation}
   R' = 0.
 \end{equation}
 The limits (\ref{RegMaxLimR}) and (\ref{RegMaxLimM})
 must hold good for all $t$ and for all $(p, q)$, so using (\ref{Edef}),
(\ref{R'RH}) \& (\ref{R'RE}) with $R > 0$, $M > 0$, $S > 0$ shows that
 \begin{eqnarray}
   && \frac{M'}{\sqrt{\epsilon + f}\;} ~~,~~~~
   \frac{f'}{\sqrt{\epsilon + f}\;} ~~,~~~~
   \frac{a'}{\sqrt{\epsilon + f}\;} ~~,~~~~
   \frac{R'}{\sqrt{\epsilon + f}\;},
   \\ \nonumber \\
   && \frac{S'}{\sqrt{\epsilon + f}\;} ~~,~~~~
   \frac{P'}{\sqrt{\epsilon + f}\;} ~~,~~~~
   \frac{Q'}{\sqrt{\epsilon + f}\;} ~~,~~~~
   \frac{E'}{\sqrt{\epsilon + f}\;}
 \end{eqnarray}
 must all have finite limits, that do not have to be zero.  Using
l'H\^{o}pital's rule, each of the above limits can be expressed in the
form
 \begin{equation}\label{limLM'}
   L_{M'} = \lim_{f \rightarrow -1}
      \frac{M'}{\sqrt{\epsilon + f}\;} =
      \frac{2 M''\sqrt{\epsilon + f}\;}{f'} =
      \frac{2 M''}{L_{f'}}.
 \end{equation}
 Thus, for $f = -1$, the above condidtions for no shell crossings in
elliptic regions should be
 re-expressed in terms of these limits.

 It is worth pointing out that $E' = 0$ at $f = -1$ does not imply the
shells near an extremum in $R$ are concentric.  It is the above limits
that determine whether there is non-concentricity at $f = -\epsilon$.

Conversely, imposing $R' = 0$ forces all of (\ref{RegMaxCond2}), if we are
to avoid shell crossings.  To obtain $f = -\epsilon$, we must impose one
further requirement for a regular extremum
 --- that no surface layers should occur at $r = r_m$.  Using the results
for the normal $n_\mu$ and the extrinsic curvature $K_{ij}$ shown in the
next section, and choosing the junction surface to be at constant
coordinate radius, $r = Z = r_m$, the non-zero components are:
 \begin{eqnarray}
   n_r & = & - \frac{(R' - R E'/E)}{\sqrt{\epsilon + f}}, \\ \nonumber \\
   K_{pp} & = & \frac{- R (\epsilon + f)^{1/2}}{E^2}, \\ \nonumber \\
   K_{qq} & = & \frac{- R (\epsilon + f)^{1/2}}{E^2}.
 \end{eqnarray}
 Now at an extremum in $R$, the factor $(R' - R E'/E)$ goes from positive
to negative (because where $(R' - RE'/E) < 0$, the
 no-shell-crossing conditions require $R' < 0$), which means that $n_\mu$
flips direction. For a boundary with no surface layer we must have $n_\mu$
pointing the same way on both sides, towards increasing $r$ say, and zero
jump in the extrinsic curvature.  So, if we cut the model at the maximum
or minimum $r_m$, and match the two halves back together, we need
 \begin{equation}
   K^+_{ij}(-n_\mu) = K^-_{ij}(+n_\mu) ~~~~\Rightarrow~~~~
   K^+_{ij} = -K^-_{ij},
 \end{equation}
 which is only possible if
 \begin{equation}
   f = -\epsilon.
 \end{equation}

 If however, $r_m$ is only a shoulder
 --- i.e. $R'(r_m) = 0$, but $R'$ has the same sign on either side, then
the normal direction does not change sign, so there is no surface layer
even if $f \neq - \epsilon$.  However $g_{rr} = L^2$ goes to zero, so it
is likely that a change of coordinates could make $|R'| > 0$.

 \subsection{Summary:~ Conditions for No Shell Crossings or Surface
Layers}
 \label{SCfNSC}

 The conditions found here are exactly those on $M$, $f$ \& $a$ for LT
models (see \cite{HelLak85} which generalises those of \cite{Bar70} for $a =
0$ LT models), with extra conditions involving $S$, $P$, $Q$ also.

 \begin{tabular}{l|l|l|l|l}
 $\epsilon$ & $R'$  & $f$      & $M',~f',~a'$ & $S',~P',~Q'$ \\
 \hline
 \hline
 $+1$       & $> 0$ & all      & $M' \geq 0$  &
      $\frac{\sqrt{(S')^2 + (P')^2 + (Q')^2}}{S} \leq \frac{M'}{3M}$ \\
 \hline
 $+1$       & $> 0$ & $\geq 0$ & $f' \geq 0$  &
      $\frac{\sqrt{(S')^2 + (P')^2 + (Q')^2}}{S} \leq \frac{f'}{2f}$ \\
            &       &          & $a' \leq 0$  &
      (no condition where $f = 0$) \\
            &       &          & but not all 3 equalities at once & \\
 \hline
 $+1$       & $> 0$ & $< 0$    & $\frac{2 \pi M}{(-f)^{3/2}} \left(
                                  \frac{M'}{M} - \frac{3 f'}{2 f} \right)
                                  + a' \geq 0$
                                              & \\
            &       &          & $a' \leq 0$  & \\
            &       &          & but not all 3 equalities at once & \\
 \hline
 \hline
 $+1$       & $= 0$ & $-1$     & $M' = 0$, $f' = 0$, $a' = 0$
                                              &
                                           $S' = 0$, $P' = 0$, $Q' = 0$ \\
            & $R'' > 0$
                    &          & ($f = -1$ for no surface layer)
                                              & \\
            & neck  &          & $\frac{2 \pi M}{(-f)^{3/2}} \left(
                                  \frac{M''}{M} - \frac{3 f''}{2 f} \right)
                                  + a'' \geq 0$
                                              &
     $\frac{\sqrt{(S'')^2 + (P'')^2 + (Q'')^2}}{S} \leq \frac{M''}{3M}$ \\
            &       &          & $a'' \leq 0$ & \\
 \hline
 $+1$       & $= 0$ & $-1$     & $M' = 0$, $f' = 0$, $a' = 0$
                                              &
                                           $S' = 0$, $P' = 0$, $Q' = 0$ \\
            & $R'' < 0$
                    &          & ($f = -1$ for no surface layer)
                                              & \\
            & belly &          & $\frac{2 \pi M}{(-f)^{3/2}} \left(
                                  \frac{M''}{M} - \frac{3 f''}{2 f} \right)
                                  + a'' \leq 0$
                                              &
    $-\frac{\sqrt{(S'')^2 + (P'')^2 + (Q'')^2}}{S} \geq \frac{M''}{3M}$ \\
            &       &          & $a'' \geq 0$ & \\
 \hline
 \hline
 $+1$       & $< 0$ & all      & $M' \leq 0$  &
     $-\frac{\sqrt{(S')^2 + (P')^2 + (Q')^2}}{S} \geq \frac{M'}{3M}$ \\
 \hline
 $+1$       & $< 0$ & $\geq 0$ & $f' \leq 0$  &
     $-\frac{\sqrt{(S')^2 + (P')^2 + (Q')^2}}{S} \geq \frac{f'}{2f}$ \\
            &       &          & $a' \geq 0$  &
      (no condition where $f = 0$) \\
            &       &          & but not all 3 equalities at once & \\
 \hline
 $+1$       & $< 0$ & $< 0$    & $\frac{2 \pi M}{(-f)^{3/2}} \left(
                                  \frac{M'}{M} - \frac{3 f'}{2 f} \right)
                                  + a' \leq 0$
                                              & \\
            &       &          & $a' \geq 0$  & \\
            &       &          & but not all 3 equalities at once & \\
 \hline
 \hline
 \end{tabular}

 \section{Impossibility of a Handle Topology}

 Since the function $E$ has the effect of making the distance between
adjacent constant $r$ shells depend on angle, this allows us to create a
wormhole that is bent, so that the two asymptotic world sheets on either
side can be thought of as intersecting in the embedding.

 This leads to the question of whether those two world sheets can be
smoothly joined across a junction surface.  In fact the possibility of
matching the two world sheets together can be considered independently of
whether there is a natural embedding that would allow them to intersect at
the appropriate angles.

\begin{figure}
 \begin{center}
 \parbox{14cm}{
 \includegraphics[scale = 0.85]{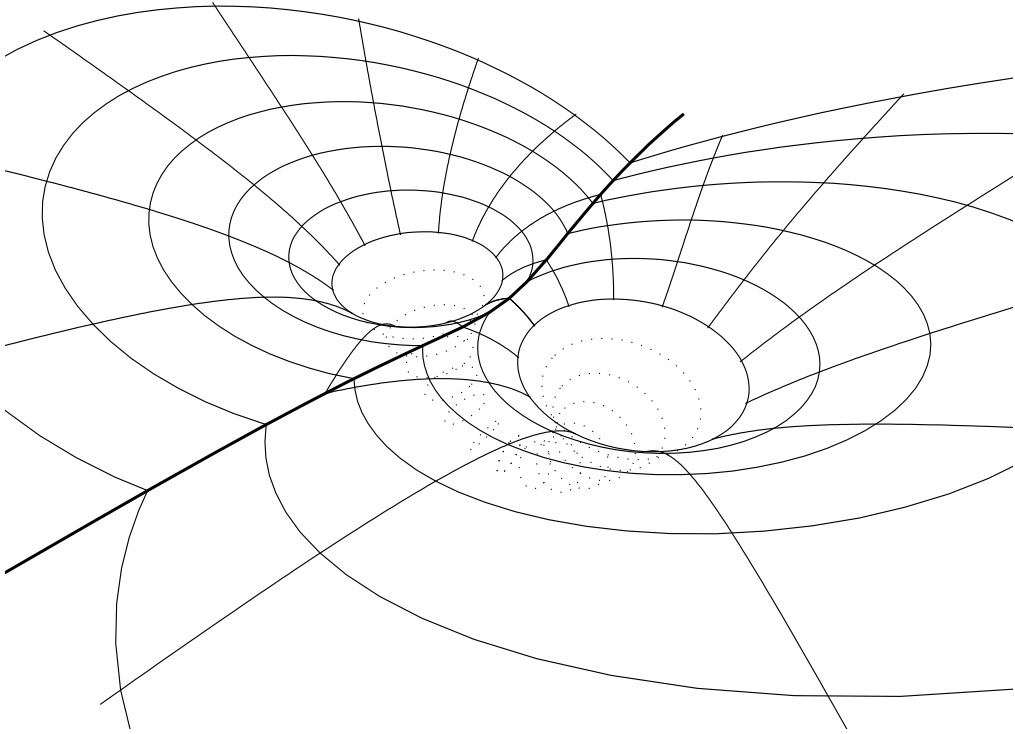}
 \caption{
 \label{Handlefig}
 \footnotesize
 Conceptual illustration of joining a Szekeres wormhole model to itself
across a boundary surface $\Sigma$, shown as a heavy line.  The boundary
may be close to the wormhole, as shown, or out in the asymptotic regions.
There is no significance to the change from solid to dotted circles, other
than picture clarity.  The handle topology is shown as an embedding of a
constant time section, with one angular coordinate suppressed, although a
valid matching across $\Sigma$ does not require the result to have a
natural embedding. However, it is shown that the matching fails because it
is not preserved by the model evolution.}
 }
 \end{center}
 \end{figure}

 Thus we investigate whether it is possible to create a Szekeres model
with a handle topology in the following way.
 Take a wormhole topology
 --- an $\epsilon = +1$ model with $r = 0$, $f(0) = -1$ at the wormhole \&
$f < 0$ nearby
 --- and let it be mirror symmetric about $r = 0$.  Choose a comoving open
surface $\Sigma$ on one side of the wormhole, and its mirror image on the other
side, and match the two sheets together along $\Sigma$, as shown schematically
in Fig. \ref{Handlefig}.  Because the 2 sheets are mirror images, this is
equivalent to matching $\Sigma$ to its own mirror image.

 ${}$ \vspace{15cm}

     To implement this we choose a comoving surface:
 \begin{equation}
   r_\Sigma = Z(p,q)
 \end{equation}
 and surface coordinates:
 \begin{equation}
   \xi^i = (t, p, q).
 \end{equation}

 The two fundamental forms and the normal are calculated in appendix
\ref{JCcalcs}.

 Obviously 1st fundamental forms match by construction, and normal vectors
are equal and opposite:
 \begin{equation}
   n_\mu^+ = - n_\mu^-.
 \end{equation}
 In fact the 2 surfaces $\Sigma^+$ \& $\Sigma_-$ are identical except for
the sign of $n_\mu$.  Thus $K^-_{ij} = - K^+_{ij}$, so the only way to
match the 2nd fundamental forms, $K^+_{ij} = K^-_{ij}$, is to make them
zero:
 \begin{equation}
   K^\pm_{ij} = 0.
 \end{equation}
 The $K_{pt} = -K_{pt}$ \& $K_{qt} = -K_{qt}$ equations give
 \begin{eqnarray}
   R' = 0 & ~~\Rightarrow~~ &
      r_\Sigma \mbox{~is at a shell crossing unless~} Z = \mbox{~const} \\
 \mbox{or~~} \dot{R} = 0 & ~~\Rightarrow~~ &
      \mbox{Static: Not possible} \\
 \mbox{or~~} Z_p = 0 ~,~~ Z_q = 0 & ~~\Rightarrow~~ &
      r_\Sigma = Z = \mbox{~constant}.
 \end{eqnarray}
 If the matching surface is at constant $r$, then only a closed torus
topology is possible.  So the answer is:~ no, a handle topology is not
possible.

     Suppose $\dot{R} = 0$ possible, then it should be possible to solve
 \begin{equation}
   K_{pp} = 0 ~~,~~~~~~ K_{pq} = 0 ~~,~~~~~~  K_{qq} = 0
 \end{equation}
 for $Z(p,q)$, by specifying suitable functions $E(r,p,q)$ \& $R(r)$ on
an initial time slice.  In other words, you can probably match on a
constant time slice, but the matching is not preserved by the model
evolution.

 \section{Szekeres Wormholes?}

It has been shown in \cite{Hel87} that LT models can describe the
Schwarzschild-Kruskal-Szekeres manifold, as well as models that have the same
topology but non-zero density.  It has also been shown that the matter flows
from past to future singularity, with possibly some matter escaping to ${\cal
J}^+$ or some being captured from ${\cal J}^-$.  The effect of the introduction
of matter on the causal structure is to split the Kruskal event horizons and
reduce communication through the wormhole.  The locus $R = 2 M$ is an apparent
horizon, but not an event horizon, and light rays fall irrevocably through the
AH towards the singularity wherever $M' > 0$. Only if the density is (locally)
zero is $R = 2M$ (locally) null.  Only if the density is everywhere zero is $R
= 2 M$ the event horizon.  (See also \cite{Bar70} for a study of light rays and
AHs in a collapsing LT model with $a = 0$.)

 Since LT models are a subset of Szekeres models, it is of interest to
look at the properties of the Szekeres generalisation, and determine how the
loss of spherical symmetry in the Szekeres model affects the LT result.

 In particular, given the anisotropy of the metric and the fact that the
proper separation of constant $r$ shells varies with $p$ \& $q$, is it
possible for null or timelike paths to pass through a neck or wormhole, by
choosing a path along which distances have been made shorter by the
particular form of $E$?  In other words, can one construct a Szekeres
wormhole that is traversible?

 For a wormhole, we require an elliptic region, in order to create a
``neck" -- a regular minimum in $R(t = const, r)$,
 \begin{equation}
   -1 \leq f < 0 ~~,~~~~~~~~ \epsilon = +1,
 \end{equation}
 but the asymptotic regions may be described by elliptic, parabolic, or
hyperbolic regions.

 \subsection{The Fastest Way Out}\label{TFWO}

 The general null condition gives
 \begin{eqnarray}
   0 = k^\alpha k^\beta g_{\alpha \beta} & = &
   (-1) \, (k^t)^2 +
   \frac{\left( R' - R \frac{E'}{E} \right)^2}{\epsilon + f} \, (k^r)^2
   + \frac{R^2}{E^2} \, ((k^p)^2 + (k^q)^2)
   \\ \nonumber \\
   \Rightarrow~~~~ &&
   \frac{\left( R' - R \frac{E'}{E} \right)^2}{\epsilon + f} \,
   \left( \frac{dr}{dt} \right)^2
   = 1 - \frac{R^2}{E^2} \, \left(
   \left( \frac{dp}{dt} \right)^2 + \left( \frac{dq}{dt} \right)^2 \right).
 \end{eqnarray}
 It is obvious that at each event $dr/dt$ is maximised by choosing $k^p =
0 = k^q$.  Since $R$ is independent of $(p, q)$, this also gives the
direction of maximum $dR/dt|_{null}$ at any event.  We will call this
``radial" motion, and radial null paths ``rays".  Thus, the DE
 \begin{equation}
   t'_n = \left. \frac{dt}{dr} \right|_n = \frac{j}{\sqrt{1 + f}\;}
      \left( R' - \frac{R E'}{E} \right) ~~,~~~~~~~~ j = \pm 1
      \label{RadialNullEq}
 \end{equation}
 in principle solves to give
 \begin{equation}
   t = t_n(r)
 \end{equation}
 along the "ray".  We don't expect this to be geodesic, but we regard it
as the limit of a sequence of accelerating timelike paths, and thus the
boundary to possible motion through a wormhole.  The acceleration of this
path may be calculated from $a^\alpha = k^\beta \nabla_\beta k^\alpha$, as
given in appendix \ref{AccelCalcs}.

 \subsection{Apparent Horizons}\label{AH}

 The areal radius along a `ray' is
 \begin{eqnarray}
   R_n & = & R(t_n(r), r), \\
   (R_n)' & = & \dot{R} \, t'_n + R'   \label{AHeq0}
   \\ \nonumber \\
   & = & j \frac{\dot{R}}{\sqrt{1 + f}\;}
      \left( R' - \frac{R E'}{E} \right) + R'   \label{AHeq}
   \\ \nonumber \\
   & = & \ell j \frac{\sqrt{\frac{2 M}{R} + f}\;}{\sqrt{1 + f}\;}
         \left( R' - \frac{R E'}{E} \right) + R' ~~,~~~~~~~~ \ell = \pm 1.
         \label{AHeq2}
 \end{eqnarray}
 These rays are momentarily stationary when
 \begin{equation}
    (R_n)' = 0.   \label{AHcondit}
 \end{equation}
 Now light rays initially along constant $p$ and $q$ will not remain so,
owing to the anisotropy of the model.  However, since these ``radial"
directions are at each point the fastest possible escape route, we define
this locus to be the apparent horizon (AH).

 (Szekeres \cite{Sze75b} defined a trapped surface as the locus where null
geodesics that are (momentarily) `radial' have zero divergence, $k^\mu{}_{;\mu}
= 0$, where $k^\mu{}_{;\nu} k^\nu = 0$, $k_\mu k^\mu = 0$, $k^p = 0 = k^q$.  He
obtained
 \begin{equation}
   k^\mu{}_{;\mu} = \frac{2}{R} \left( R' - \frac{R E'}{E} \right)
   \left( \frac{\dot{R}}{\sqrt{1 + f}\;} + j \right).
 \end{equation}
 Given the anisotropy of the model, one doesn't expect this to be the same
locus as our AH.)

 Assuming a normal spacetime point will have
 non-zero metric components, and taking $R$ increasing with $r$ on
constant $t$ slices,
 \begin{equation}
   R' > 0 ~~~~~~\mbox{and}~~~~~~ \left( R' - \frac{R E'}{E} \right) > 0,
 \end{equation}
 we require
 \begin{equation}
   \ell j = -1,
 \end{equation}
 i.e.
 \begin{eqnarray}
   \mbox{Either~~(future AH:~~AH$^+$)}
   && j = +1 ~~~~\mbox{(outgoing rays)}   \nonumber \\
   && \ell = -1 ~~~~\mbox{(in a collapsing phase)} \\
   \mbox{Or~~(past AH:~~AH$^-$)}
   && j = -1 ~~~~\mbox{(incoming rays)}   \nonumber \\
   && \ell = +1 ~~~~\mbox{(in an expanding phase)}.
 \end{eqnarray}
 Note that we want `outgoing' to mean moving away from the neck at $r = 0$.  A
ray passing through the neck would change from incoming to outgoing at $r
= 0$, and, since $R'$ flips sign there, $j$ would also have to flip there.

 \subsubsection{The Apparent Horizon and its Location with Respect to
                $E' = 0$.}

 Define
 \begin{equation}\label{defDAH}
 D := \sqrt{1 + f} - \sqrt{\frac{2M}R + f}.
 \end{equation}
 Then
 \begin{equation}\label{AHequiv}
 (D > 0) \Longleftrightarrow (R > 2M).
 \end{equation}
 Since $M/R$ and $(2M/R + f)$ are positive, we see that $D \geq 1$ leads to a
contradiction, and so
 \begin{equation}\label{D<1}
 D < 1.
 \end{equation}
 However, $|D|$ can be greater than 1 because $D < -1$ is not prohibited.  We
have
 \begin{equation}\label{Dlthan-1}
 (D < -1) \Longrightarrow \left(R < \frac {M}{1 + \sqrt{1 +
f}}\right).
 \end{equation}
 This will always occur when $R$ is close to the Big Bang/Big
Crunch.

Using $D$, the equation of the AH is
 \begin{equation}\label{eqAH}
 RE' + DR'E = 0,
 \end{equation}
 and in terms of $p$ and $q$ this equation is
 \begin{equation}\label{eqAHpq}
 \left(\frac{S'}S - D\frac{R'}R\right)\left[(p - P)^2 + (q -
Q)^2\right] + 2[(p - P)P' + (q - Q)Q'] -  S^2\left(\frac {S'}S + D
\frac{R'}R\right) = 0.
 \end{equation}
 The discriminant of this with respect to $p$ is
 \begin{equation}\label{discpAH}
 \Delta_p = 4P'^2 - 4\left(\frac{S'}S - D\frac{R'}R\right)
\left[\left(\frac{S'}S - D\frac{R'}R\right)(q - Q)^2 + 2 (q - Q)Q' -  S^2
\left(\frac{S'}{S} + D\frac{R'}{R}\right)\right].
 \end{equation}
 The discriminant of this with respect to $q$ is
 \begin{equation}\label{discqAH}
 \Delta_q = 64\left(\frac{S'}S - D\frac{R'}R\right)^2 \left[P'^2 +
Q'^2 +  S^2 \left(\frac{S'^2}{S^2} - D^2 \frac {R'^2} {R^2}\right)\right].
 \end{equation}

 Now, if $\Delta_q < 0$ everywhere, then $\Delta_p < 0$ for all
$q$, in which case there is no $p$ obeying (\ref{eqAHpq}), i.e.
the apparent horizon does not intersect this particular surface of
constant $(t, r)$.

 If $\Delta_q = 0$, then $\Delta_p < 0$ for all $q$ except one value $q = q_0$,
at which $\Delta_p = 0$.  At this value of $p = p_0$, (\ref{eqAHpq}) has a
solution, and so the intersection of the apparent horizon with this one
constant $(t, r)$ surface is a single point.  Note that the situation when the
apparent horizon touches the whole 3-dimensional $t =$ const hypersurface at a
certain value of $t$ is exceptional, this requires, from (\ref{eqAHpq}), that
$P' = Q' = S' = R' = 0$ at this value of $t$.  The first three functions being
zero mean just spherical symmetry, but the fourth one defines a special
location, as mentioned at the beginning of sec. \ref{OaPoSCiaSoCtar}.  These
equations hold in the Datt-Ruban \cite{Datt1938, Ruba1968, Ruba1969} solution.

 If $\Delta_q > 0$, then $\Delta_p > 0$ for every $q$ such that
$q_1 < q < q_2$, where
 \begin{equation}\label{solqAH}
q_{1,2} = \frac {- Q' \pm \sqrt{P'^2 + Q'^2 +  S^2 \left(\frac{S'^2}{S^2} -
D^2\frac{R'^2}{R^2}\right)}} {S'/S - DR'/R}
 \end{equation}
 and then a solution of (\ref{eqAHpq}) exists given by
 \begin{equation}\label{solpAH}
p_{1,2} = \frac {- P' \pm \sqrt{- \left[\left(\frac{S'}{S} -
D\frac{R'}{R}\right)(q - Q) + Q'\right]^2 + P'^2 + Q'^2 + S^2
\left(\frac{S'^2}{S^2} - D^2\frac{R'^2}{R^2}\right)}} {S'/S - DR'/R}.
 \end{equation}
 Except for the special case when $S'/S = DR'/R$, these values lie on a circle
in the $(p, q)$ plane, with the center at
 \begin{equation}\label{cenAH}
\left(p_{AH}, q_{AH}\right) = \left(P - \frac {P'} {S'/S - DR'/R},
Q - \frac {Q'} {S'/S - DR'/R}\right),
 \end{equation}
 and with the radius $L_{AH}$ given by
 \begin{equation}\label{radAH}
{L_{AH}}^2 = \frac {P'^2 + Q'^2 +  S^2 \left(\frac{S'^2}{S^2} -
D^2\frac{R'^2}{R^2}\right)} {(S'/S - DR'/R)^2} .
 \end{equation}
The special case $S'/S = DR'/R$ (when  the locus of AH in the $(p, q)$ plane is
a straight line) is again an artefact of the Riemann projection because this
straight line is an image of a circle on the sphere.

 In summary, the intersection of AH with the $(p, q)$-plane is
 \begin{itemize}
 \item   nonexistent when $R'^2 /R^2 >  \Phi^2/D^2$ (this is the
same $\Phi$ as for the shell crossing);
 \item   a single point when $R'^2 /R^2 =  \Phi^2/D^2$;
 \item   a circle or a straight line when $R'^2 /R^2 <
\Phi^2/D^2$.
 \end{itemize}
 The condition $R'^2 /R^2 <  \Phi^2/D^2$ is consistent with the
condition for no shell crossings, eq. (\ref{noshcr}), when $|D| < 1$.  We
already know that necessarily $D < 1$, but $D < -1$ is not excluded.

 With $|D| < 1$, when the intersection of AH with $(t = {\rm
const}, r = {\rm const})$ is a single point, a shell crossing is
automatically excluded.

 Note that from (\ref{eqAH}) and from the assumptions $R > 0$, $E >
0$ and $R' > 0$ we have
 \begin{equation}\label{signDE}
(D > 0) \Longrightarrow (E' < 0)
 $$ $$
 (D < 0) \Longrightarrow (E'
> 0).
 \end{equation}
 But $D > 0$ and $D < 0$ define regions independent of $p$ and $q$.  Hence, on
that surface, on which $D > 0$, $E' < 0$ on the whole of AH.  Where $D < 0$, $E'
> 0$ on the whole of AH.  This implies that the $E' = 0$ circle and the
AH cannot intersect unless they coincide.  Indeed, these circles lie in
parallel planes, by the same argument that was used at the end of sec.
\ref{OaPoSCiaSoCtar}: the line on the $(t, r) =$ const surface defined by
(\ref{eqAH}) has the property $E'/E = - DR'/R =$ const, and so it must be a
circle in a plane parallel to the $E' = 0$ great circle.  It follows that of
the three circles ($E' = 0$, SC and AH), no two can intersect unless they
coincide.

\begin{figure}
 ${}$ \hspace{-15mm}
 \begin{center}
 \parbox{14cm}{
 \includegraphics[scale = 0.75]{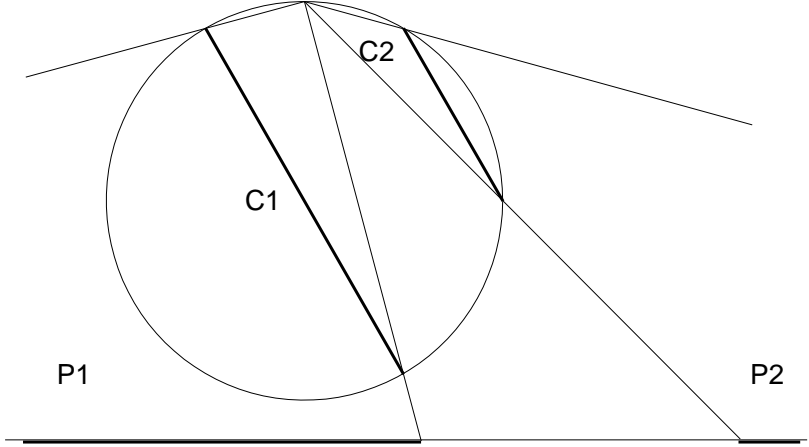}
 \caption{
 \label{proj1fig}
 \footnotesize
 The circles C1 and C2 on a sphere (seen here edge on) will project onto
the plane (seen here as the horizontal line) as the circles P1 and P2 that are
outside each other.  Only parts of P1 and P2 are shown here.  Circle C1 is the
$E' = 0$ set, circle C2 is the apparent horizon circle.
 }
 }
 \end{center}
 \end{figure}

\begin{figure}
 ${}$ \hspace{-15mm}
 \begin{center}
 \parbox{14cm}{
 \includegraphics[scale = 0.75]{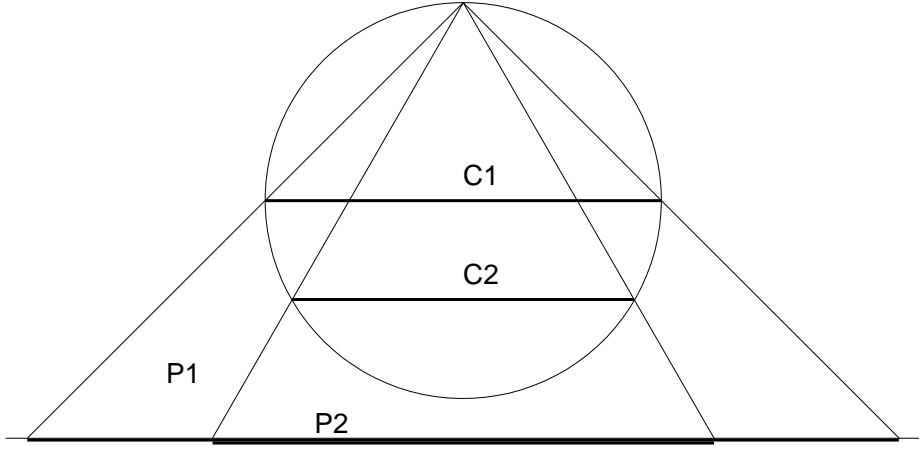}
 \caption{
 \label{proj2fig}
 \footnotesize
 The same circles as in Fig.
\ref{proj1fig} projected onto a plane from a different pole will project as one
inside the other.  The transition from the situation of Fig. \ref{proj1fig} to
that of Fig. \ref{proj2fig} is continuous and occurs when the sphere is
rotated, but the pole and the plane are not moved. Then one of the circles (C1
when a clockwise rotation is applied to Fig. \ref{proj1fig}) will pass through
the pole at one value $\varphi = \varphi_0$ of the rotation angle. Its image on
the plane is acquiring a larger and larger radius with increasing $\varphi$,
until it becomes a straight line when $\varphi = \varphi_0$.  When $\varphi$
increases further, the straight line bends in the opposite direction so that it
surrounds the second circle P2.
 }
 }
 \end{center}
 \end{figure}

 When the $E' = 0$ and AH circles are disjoint, they may either be one inside
the other or each one outside the other. However, when projected back onto the
sphere, these two situations turn out to be topologically equivalent: depending
on the position of the point of projection, the same two circles may project
onto the plane either as one circle inside the other or as two separate
circles, see Figs. \ref{proj1fig} and \ref{proj2fig}.

 \bigskip

 \bigskip

 \subsubsection{Location of the AH Compared with $R = 2M$}

 Along $R = 2M$
 \begin{equation}
   (R_n)' = R' (1 + \ell j) - \frac{R E'}{E},
 \end{equation}
 so $R = 2M$ is not the AH except where $E' = 0$.

 Eq (\ref{AHeq2}) with (\ref{AHcondit}) and $\ell j = -1$ can be written
 \begin{eqnarray}
   R_{AH} & = & \frac{2 M (1 - R E'/R' E)^2}
      {1 + f [1 - (1 - R E'/R' E)^2]}
   \\ \nonumber \\
   & = & \frac{2 M (1 - V)^2}{1 + f (2 V - V^2)}
      ~~,~~~~~~~~ V = R E'/R' E.   \label{AHeqV}
 \end{eqnarray}

 The effect of $E(r,p,q)$ is to create a dipole in the geometry and
density around each $(t,r)$ shell, with $E' = 0$ on an ``equator", and
extreme values
 \begin{equation}
   \left. \frac{E'}{E} \right|_{\rm extreme} =
      \pm \frac{\sqrt{(S')^2 + (P')^2 + (Q')^2}\;}{S}
 \end{equation}
 at the poles.

 It is clear then that ``radial" displacements between two nearby surfaces
of constant $r$ are shortest where $E'/E$ is maximum, and light rays move
outwards fastest (max $dr/dt$, min $dt/dr$, i.e. most rapid transfer between
constant $r$ shells at the same $(p,q)$ value).  It has also been shown that
the density is minimum here.  The longest ``radial" displacements, slowest
light ray motion and maximum density occur at the opposite pole.

 We will call the direction where $E'/E$ is maximum, the `fast' pole,
and where $E'/E$ is minimum, the `slow' pole.

 Now the conditions for no shell crossings require
 \begin{equation}
   \left. \frac{E'}{E} \right|_{\rm extreme} < \frac{M'}{3M}
 \end{equation}
 and for an elliptic region we have
 \begin{equation}
   \frac{M'}{3M} < \frac{R'}{R},
 \end{equation}
 so
 \begin{equation}
   V^2 < 1
   ~~,~~~~~~~~
   (1 - V)^2 > 0
   ~~,~~~~~~~~
   -3 < (2 V - V^2) < 1.
 \end{equation}

In those places on the AH, where $V = 0 = E'$, we see that the surface $R = 2M$
intersects the AH at all times, but the AH is a kind of oval with half inside
$R = 2M$ and half outside.

 For $f = -1$
 \begin{equation}
   \frac{R_{AH}}{2M} = 1
 \end{equation}
 regardless of $V$.  So it is clear that AH$^+$ \& AH$^-$ cross in a
 2-sphere at the neck of the wormhole ($f = -1$) at the moment of maximum
expansion ($R = 2M$), as in LT.  Note also that at the bang, wherever $a'
\neq 0$, $R \rightarrow 0$ and $R' \rightarrow \infty$ imply $V
\rightarrow 0$, and the anisotropy becomes negligible.  Similarly for the
crunch.

 But in general, for all $0 \geq f > -1$
 \begin{equation}
   \frac{4}{1 + 3 |f|} \geq \frac{R_{AH}}{2M} \geq 0
 \end{equation}
 and $R_{AH}/2M$ decreases monotonically as $V$ goes from $-1$ to $1$.
Note that $|V_{\rm extreme}|$ is likely to be less than $1$, and also that
the maximum \& minimum values of $R_{AH}$ do not have a simple
relationship.

 We have that $R_{AH}/2M < 1$ where $V > 0$, i.e. where $E' > 0$.  In
other words, taking a $(t,r)$ shell that intersects the AH at the fast
pole, the light rays move fastest between the shells exactly where the
shell is just emerging from the AH.

 Some other features of the AH locus are discussed in appendix
\ref{AHfeatures}.

 \subsection{Causal Structure of A Szekeres Wormhole}\label{CSoASW}

 We shall next establish whether a radial null ray can pass through a
Szekeres wormhole.  We shall have to treat the neck separately from every
other $r$ value, because of the need to treat the $f \rightarrow -1$ limit
carefully.

 \subsubsection{Can $E' > 0$ Compensate for $M' > 0$?}

 As noted in \cite{Hel87} the introduction of matter into a wormhole slows
the progress of light rays through it.  Can this effect be compensated for
by a suitable choice of $E' > 0$?  Since the vacuum case is $M' = 0 = E'$,
for which we know the behaviour, we are only interested in the effects of
varying $M$ and $E$.

 We start with the gradient of the null rays, eq (\ref{RadialNullEq})
 and use the substitutions (\ref{ellevRS}) and (\ref{R'RE}) with
(\ref{phi1phi2}) for $R$, \& $R'$ in terms of $\eta$ \& $r$, but we note
that, if we choose the future AH
 --- i.e. outgoing rays in a collapsing region
 --- then
 \begin{equation}
   j = +1 ~~~~\mbox{and}~~~~ \pi < \eta \leq 2 \pi.   \label{AH+jeta}
 \end{equation}
 So the gradient of the null rays in terms of $\eta$ \& $r$ is
 \begin{eqnarray}
      \left. \frac{dt}{dr} \right|_n  & = &
      \frac 1{\sqrt{1+f}}\left( \frac{M ( 1 - \cos \eta)}{(-f) \sqrt{1 + f)}\;} \right)
      \left\{
      - \left(
      \frac{\sin \eta}{(1 - \cos \eta)^2}
      \right)
      \frac{(-f)^{3/2} a'}{M}
      \right.
      \nonumber \\ \nonumber \\
   && - \left(
      1 - \frac{3 \sin \eta (\eta - \sin \eta)}{2 (1 - \cos \eta)^2}
      \right)
      \frac{f'}{f}
      \nonumber \\ \nonumber \\
   && \left.
      + \left(
      1 - \frac{\sin \eta (\eta - \sin \eta)}{(1 - \cos \eta)^2}
      \right)
      \frac{M'}{M}
      - \frac{E'}{E}
      \right\}.
 \end{eqnarray}
 Consider a region in which $R' > 0$ and $M' > 0$.  Now since, in the
above range of $\eta$
 \begin{eqnarray}
   2 \geq 1 - \cos \eta \geq 0
   ~~~~&,&~~~~
   0 \geq \frac{\sin \eta}{(1 - \cos \eta)} \geq -\infty,
   \nonumber \\ \nonumber \\
   1 \leq
   1 - \frac{\sin \eta (\eta - \sin \eta)}{(1 - \cos \eta)^2}
   \leq \infty
   ~~~~&,&~~~~
   1 \leq
   1 - \frac{3 \sin \eta (\eta - \sin \eta)}{2 (1 - \cos \eta)^2}
   \leq \infty,
 \end{eqnarray}
 the coefficient of $M'/M$ is always positive, and the coefficient of
$E'/E$ is always negative.  In particular, because of the no shell
crossing condition (\ref{MrCond}), $|E'/E| \leq M'/3M$, the $E'/E$ term
gives at most a partial cancellation of the $M'/M$ term.  Thus it is
evident that varying $E'/E$ cannot compensate for the effect of
 non-zero $M'$ on the gradient of the radial rays.

 We turn to the AH equation (\ref{AHcondit}) with (\ref{AHeq})
and(\ref{AH+jeta}).  We find the future AH equation in terms of $\eta$ and
$r$ may be written
 \begin{eqnarray}
   0 & = & \left[
      1 + \sqrt{\frac{(-f)}{(1 + f)}}\; \frac{\sin \eta}{(1 - \cos \eta)}
      \right]
      \Bigg\{- \frac{(-f)^{3/2} a'}{M}
      \left(
      \frac{\sin \eta}{(1 - \cos \eta)^2}
      \right)
      \nonumber \\ \nonumber \\
   && - \frac{f'}{f}
      \left(
      1 - \frac{3 \sin \eta (\eta - \sin \eta)}{2 (1 - \cos \eta)^2}
      \right) + \frac{M'}{M}
      \left(
      1 - \frac{\sin \eta (\eta - \sin \eta)}{(1 - \cos \eta)^2}
      \right)
      \Bigg\}
      \nonumber \\ \nonumber \\
   && - \left[
      \sqrt{\frac{(-f)}{(1 + f)}}\; \frac{\sin \eta}{(1 - \cos \eta)}
      \right]
      \frac{E'}{E}.
   \label{etaAHeq}
 \end{eqnarray}
 The solution is the parametric locus $\eta = \eta_{AH}(r)$.  It is
evident that if $E' = 0$, varying $M'$ has no effect at all on the AH
locus for a given $M$, as the solution is
 \begin{equation}
   \sqrt{\frac{(-f)}{(1 + f)}}\; \frac{\sin \eta}{(1 - \cos \eta)} = -1
   ~~~~\Rightarrow~~~~
   \cos \eta = 1 + 2 f
   ~~~~\Rightarrow~~~~
   R = 2 M.
 \end{equation}
 Similarly $E'$ has no effect when $\eta = \pi$, and (see appendix
\ref{AHfeatures}) when $\eta = 0$ or $2 \pi$.  On the other hand, the
effect of varying $E'$ is influenced by the value of $M'$. Analysing the
slope of this curve leads to pretty daunting expressions, but is
fortunately not necessary.

 Consider now the slope of a surface $R(t,r) = \alpha M(r)$ in a
collapsing region
 \begin{equation}
   R = \alpha M
   ~~~~\Rightarrow~~~~
   \dot{R}^2 = \frac{2 M}{\alpha M} + f
   ~~~~\Rightarrow~~~~
   \left. \frac{dt}{dr} \right|_{R = \alpha M} =
      \frac{R' - \alpha M'}{\sqrt{\frac{2}{\alpha} + f}\;},
 \end{equation}
 where $\alpha > 0$.  This is null or outgoing timelike wherever
 \begin{equation}
   \frac{(R' - \alpha M E' / E)}{\sqrt{1 + f}\;} \leq
      \frac{R' - \alpha M'}{\sqrt{\frac{2}{\alpha} + f}\;}.
 \end{equation}
For $M' = 0$, which forces $E' = 0$ (by eq. (\ref{MrCond})), the equality
obviously requires $\alpha = 2$, giving the event horizon in a vacuum model,
and all $R = \alpha M$ surfaces are outgoing timelike for $\alpha > 2$, and
spacelike for $\alpha < 2$.  For $M' > 0$, the condition $E'/E < M'/3M$ ensures
the numerator of the lhs is no less than
 \begin{equation}
   R' - \alpha M' / 3.
 \end{equation}
 For any given $M' > 0$, $R' > 0$ and $f > -1$, this is always greater
than the numerator on the rhs, so, to satisfy the equality, the
denominator on the lhs must be greater than that on the rhs, so once again
 \begin{equation}
   \alpha > 2.
 \end{equation}
 Thus $R = \alpha M$ surfaces can only be tangent to outgoing null rays
for $R = \alpha M > 2 M$, and for $R < 2 M$ they are spacelike, incoming
null, or incoming timelike.

 This allows us to conclude that, along the entire length of the future $R
= 2 M$ surface, outgoing rays pass inside it, or run along it where $M' =
0$.  By (\ref{ellevRS}), the maximum $R$ along any given constant $r$
worldline in an elliptic region is at $\eta = \pi$, when
 \begin{equation}
   R_{max} = \frac{2 M}{(-f)} \geq 2 M,
 \end{equation}
 while $R$ grows without bound in parabolic and hyperbolic regions.  Thus
every particle worldline encounters the future $R = 2M$ surface (and the
past surface), leaving no room for any rays that arrive at the future $R
= 2 M$ surface to escape to ${\cal J}^+$.

 The time reverse of these arguments applies to the past AH, which lies
in an expanding region and has incoming rays running along it or passing
out of it.

 To complete the argument, we must consider the limits at the neck, $f
\rightarrow -1$ where several derivatives are zero.

 \subsubsection{The AH at the Neck}

 Now we turn to consider the AH at the neck.

 The differential of (\ref{AHeq}) with (\ref{AHcondit}) and $dp = 0 = dq$
gives us
 \begin{eqnarray}
   0 & = & j \frac{\dot{R}}{\sqrt{1 + f}\;}
      \left( R' - \frac{R E'}{E} \right) \Bigg\{
         \frac{\ddot{R} \, dt + \dot{R}' \, dr}{\dot{R}}
         - \frac{f' \, dr}{2(1 + f)}
   \nonumber \\ \nonumber \\
   &&    + \frac{\left( \dot{R}' \, dt + R'' \, dr
            - \frac{(\dot{R} \, dt + R' \, dr) E'}{E}
            - \frac{R E'' \, dr}{E} + \frac{R (E')^2 \, dr}{E^2} \right)}
            {\left( R' - \frac{R E'}{E} \right)} \Bigg\}
   \nonumber \\ \nonumber \\
   &&    + \dot{R}' \, dt + R'' \, dr,
   \\ \nonumber \\
   \left. \frac{dt}{dr} \right|_{AH} & = & - \left\{
   j \frac{\dot{R}'}{\sqrt{1 + f}\;} \left( R' - \frac{R E'}{E} \right)
   - j \frac{\dot{R} f'}{2 (1 + f)^{3/2}}
           \left( R' - \frac{R E'}{E} \right) \right.
   \nonumber \\ \nonumber \\
   && \left.
      + j \frac{\dot{R}}{\sqrt{1 + f}\;} \left( R'' - \frac{R' E'}{E}
            - \frac{R E''}{E} + \frac{R (E')^2}{E^2} \right)
      + R'' \right\} \Bigg/
   \nonumber  \\ \nonumber \\
   && ~~~~\left[
   j \frac{\ddot{R}}{\sqrt{1 + f}\;} \left( R' - \frac{R E'}{E} \right)
   + j \frac{\dot{R}}{\sqrt{1 + f}\;}
         \left( \dot{R}' - \frac{\dot{R} E'}{E} \right)
   + \dot{R}' \right].
 \end{eqnarray}


At the neck of the wormhole, $r = r_n$, the regularity conditions of sec.
\ref{Regmaxmin} give us the following limits, where $L_{f'}$ etc are being
defined in each case:
 \begin{eqnarray}
   &&~~~~ f \rightarrow -1, \\
   f' \rightarrow 0, ~~~~&&~~~~
      f'/\sqrt{1 + f}\; \rightarrow L_{f'} = 2 f'' / L_{f'} \geq 0
      ~~~~\Rightarrow~~~~ (L_{f'})^2 = 2 f'', \label{199}\\
   E' \rightarrow 0, ~~~~&&~~~~
      E'/\sqrt{1 + f}\; \rightarrow L_{E'} = 2 E'' / L_{f'} \geq 0, \\
   R' \rightarrow 0, ~~~~&&~~~~
      R'/\sqrt{1 + f}\; \rightarrow L_{R'} = 2 R'' / L_{f'} > 0, \\
   \dot{R}' \rightarrow 0, ~~~~&&~~~~
      \dot{R}'/\sqrt{1 + f}\; \rightarrow L_{\dot{R}'} =
      2 \dot{R}'' / L_{f'} = (\partial / \partial t) L_{R'}.
 \end{eqnarray}
 Thus the terms ~$\dot{R}'R'/\sqrt{1 + f}\;$~, ~$\dot{R}'RE'/E\sqrt{1 +
f}\;$~, ~$R'E'/E\sqrt{1 + f}\;$~, ~$R(E')^2/E^2\sqrt{1 + f}\;$~and ~$\dot{R}'$
go to zero and the remaining numerator terms involving $\dot{R}$ cancel, down
to
 \begin{eqnarray}
   \left. \frac{dt}{dr} \right|_{AH, \rm N} & = &
   - j R'' \Bigg/ \Bigg[
   \frac{2 \ddot{R}}{L_{f'}} \left( R'' - \frac{R E''}{E} \right) +
   \frac{\dot{R}}{\sqrt{1 + f}\;}
   \left( \dot{R}' - \frac{\dot{R} E'}{E} \right)
   \Bigg], \\   \nonumber   \\
   \left. \frac{dt}{dr} \right|_{n,\rm N}
      & = & j \frac{2 (R'' - R E'' / E)}{L_{f'}}.
 \end{eqnarray}
 Since the AH only intersects the neck when $\dot{R} = 0$, the behaviour
of $\dot{R}/\sqrt{1 + f}\;$ and $\dot{R}^2/\sqrt{1 + f}\;$ must still be
determined (and that of $\dot{R}'$ will be verified).


 At the moment of maximum expansion in the neck we have
 \begin{eqnarray}
   && R = 2 M ~~,~~~~~~
   \dot{R} = 0 ~~,~~~~~~
   \sqrt{1 + f} = 0,   \\
   && \eta = \pi ~~,~~~~~~
   \cos \eta = -1 ~~,~~~~~~
   \sin \eta = 0,
 \end{eqnarray}
 and so, using
 \begin{eqnarray}
   {}\!\!\!\!\!\!\!\!\hspace*{-2cm}                      
   R' & = & \frac{M}{(-f)} \left\{
      \left( \frac{M'}{M} - \frac{f'}{f} \right) (1 - \cos \eta)
      - \left( \frac{M'}{M} - \frac{3 f'}{2 f} \right)
      \frac{\sin \eta (\eta - \sin \eta)}{(1 - \cos \eta)} \right\}
      \nonumber \\ \nonumber \\
   && - a' \sqrt{-f}\; \frac{\sin \eta}{(1 - \cos \eta)},
      \\ \nonumber \\
   {}\!\!\!\!\!\!\!\!
   R'' & = &
   \frac{-1}{(1 - \cos \eta)^2} \frac{M}{(-f)} \left\{
      (\eta - \sin \eta) \left( \frac{M'}{M} - \frac{3 f'}{2 f} \right)
      + \frac{a' (-f)^{3/2}}{M} \right\}^2
      \nonumber \\ \nonumber \\
   && +   \frac{\sin \eta}{(1 - \cos \eta)} \left\{
      (\eta - \sin \eta) \frac{M}{(-f)}
      \left[ \frac{2 f'}{f} \left( \frac{M'}{M} - \frac{9 f'}{8 f} \right)
      - \left( \frac{M''}{M} - \frac{3 f''}{2 f} \right) \right] \right.
      \nonumber \\ \nonumber \\
   && \left. - \sqrt{-f}\; \left( a'' + a' \frac{f'}{f} \right) \right\}
      +   (1 - \cos \eta) \frac{M}{(-f)} \left\{
      \frac{M''}{M} - \frac{f''}{f} - \frac{2 f'}{f}
      \left( \frac{M'}{M} - \frac{f'}{f} \right) \right\}
 \end{eqnarray}
 gives, by virtue of (\ref{199}) and (\ref{limLM'}):
 \begin{equation}
   L_{R'} = 2 (L_{M'} + M L_{f'}) = 4 (M'' + M f'') / L_{f'} ~~,~~~~~~
   R'' = 2 ( M'' + M f'').   \label{RrRrrMEN}
 \end{equation}
 We find the limit of $\dot{R}/\sqrt{1 + f}$ at this point by combining
(\ref{AHeq}) and (\ref{AHcondit}), to obtain
 \begin{equation}
   \left. \frac{\dot{R}}{\sqrt{1 + f}\;} \right|_{\rm MEN} =
      \frac{- j L_{R'}}{L_{R'} - R L_{E'} / E} =
      \frac{- j (M'' + M f'')}{M'' + M f'' - M E''/E},
 \end{equation}
 so it is clear that $\dot{R}^2 E'/\sqrt{1 + f}\; = 0$.  To check the limit
of $\dot{R}'$, the $r$ derivative of (\ref{RdotSq}) gives
 \begin{equation}
   \dot{R}' = \frac{1}{\dot{R}} \left( \frac{M'}{R} - \frac{M R'}{R^2}
              + \frac{f'}{2} \right)
   ~~~~\rightarrow~~~~
   \frac{\dot{R}\dot{R}'}{\sqrt{1 + f}\;} = \left( \frac{L_{M'}}{R}
              - \frac{M L_{R'}}{R^2} + \frac{L_{f'}}{2} \right),
 \end{equation}
 and because of (\ref{RrRrrMEN}) and $R = 2M$ all terms in the bracket cancel,
verifying that $\dot{R}\dot{R}'/\sqrt{1 + f}\; = 0$.

 These together with (\ref{Rddot}) give us
 \begin{eqnarray}
   \left. \frac{dt}{dr} \right|_{AH, \rm MEN} & = &
      j \frac{4 M f'' (M'' + M f'')}{L_{f'} (M'' + M f'' - M E'' / E)},
   \\ \nonumber \\
   \left. \frac{dt}{dr} \right|_{n,\rm MEN} & = &
      j \frac{4 (M'' + M f'' - M E'' / E)}{L_{f'}}.
 \end{eqnarray}


 For a light ray to pass through the neck at the moment of maximum
expansion without falling inside the AH, we need $dt/dr|_{AH,\rm MEN} >
dt/dr|_{n,\rm MEN}$, in other words
 \begin{equation}
   \frac{dt/dr|_{AH,\rm MEN}}{dt/dr|_{n,\rm MEN}} =
   \left. \frac{M f'' (M'' + M f'')}
      {(M'' + M f'' - M E'' / E)^2} \right|_{\rm MEN} > 1,
      \label{AH_n_SlopeRatio}
 \end{equation}
 or
 \begin{equation}
   \frac{(M'' + M f'') - \sqrt{M f'' (M'' + M f'')}\;}{M} <
   \frac{E''}{E} < \frac{(M'' + M f'') + \sqrt{M f'' (M'' + M f'')}\;}{M}.
   \label{ThroughCondit}
 \end{equation}
 Since $M(r)$ is positive, and both $M(r)$ and $f(r)$ are at a minimum at
the neck, i.e. $M'' > 0$ \& $f'' > 0$, we have $\sqrt{M f'' (M'' + M
f'')}\; < M'' + M f''$, and so both upper and lower limits are real and
positive.

 Can this requirement be satisfied without creating shell crossings?  The
only relevant condition is the one for $\epsilon = +1$, $R' = 0$, $f = -
1$, $R'' > 0$
 \begin{equation}
   \left. \frac{E''}{E} \right|_{\rm max} \leq \frac{M''}{3 M}.
 \end{equation}
 To be able to satisfy this as well as (\ref{ThroughCondit}) we would need
 \begin{equation}
   \frac{M'' + M f'' - \sqrt{M f'' (M'' + M f'')}\;}{M} < \frac{M''}{3 M},
 \end{equation}
 but this leads to
 \begin{equation}
   M'' ( 4 M''+ 3 M f'') < 0,
 \end{equation}
 which is clearly not possible.  Indeed, although $\rho$ is zero rather
than divergent where $E'/E = M'/3M$, where $E'/E$ exceeds $M'/3M$, the
density is negative at all times.

 Putting the maximum value, $E''/E = M''/3M$ into (\ref{AH_n_SlopeRatio})
gives
 \begin{equation}
   \left. \frac{dt/dr|_{AH,\rm MEN}}{dt/dr|_{n,\rm MEN}} \right|_{\rm max}
   = \frac{9 M f'' (M'' + M f'')}
      {(2 M'' + 3 M f'')^2},
 \end{equation}
 which rises from $0$ at $f''/M'' = 0$, and asymptotically approaches $1$
as $f''/M'' \rightarrow \infty$, i.e. vacuum.

 Therefore, even at the neck, $E' > 0$ cannot compensate for $M' > 0$, and
all rays passing through this event remain within $R \leq 2 M$, passing
from inside AH$^-$ to inside AH$^+$.

 \subsubsection{Summary}

 In a Szekeres wormhole, every particle worldline encounters $R = 2 M$,
twice for most $r$ values and once where $f = -1$, making this a pair of
 3-surfaces that span the spacetime.  The apparent horizons coincide with
$R = 2M$ at an extremum of $R(t=const, r)$
 --- a neck or belly
 --- where $f = -1$.  Where $M' = 0$ (vacuum), the $R = 2 M$ surfaces are
(locally) null.

 Assuming there is matter ($M' > 0$) somewhere within the elliptic region
describing the neck, and assuming the two regions, $r \rightarrow \pm
\infty$, are asymptotically flat, i.e. $M \rightarrow M_{tot} = $~constant
($E' \rightarrow 0$), then the event horizon is the set of rays that are
asymptotic to $R = 2 M$, but always lie outside.  The future event horizon
EH$^+$ emerges from $R = 2M$ surface, and vice-versa for EH$^-$.

 Thus we conclude that the causal structure of a regular Szekeres wormhole
is only a quantitative modification of the LT wormhole (dense black hole),
and the possible causal diagrams for Szekeres models are essentially the
same as those for LT models, as given in \cite{Hel87}.

 \subsubsection{Numerical Examples}

 A few numerical examples were produced as follows.

 We choose the 3 LT arbitrary functions to produce a Kruskal-like topology,
with the neck at $r = 0$, that is mirror symmetric about $r = 0$ and $t =
0$.  The choice must therefore satisfy $f(0) = -1$, $f'(0) = 0$, $f''(0) >
0$, $M'(0) = 0$, $M''(0) > 0$, $a(r) = -b(r)$;
 \begin{eqnarray}
   M & = & M_0 (1 + M_1 r^2)^3 ~~,~~~~~~~~ M_0, M_1 > 0, \\
   f & = & -{\rm exp}(-r^2/r_s) ~~,~~~~~~~~ r_s > 0, \\
   a & = & - \pi M / (-f)^{3/2}.
 \end{eqnarray}
 We want to choose $E$ to maximise the effect of $E' \neq 0$ along one
particular radial path.  By setting
 \begin{equation}
   P(r) = 0 = Q(r),
 \end{equation}
 so that (\ref{Edef}) is
 \begin{equation}
   E = \frac{S}{2} \left( \frac{p^2}{S^2} + \frac{q^2}{S^2} + 1 \right),
 \end{equation}
 the maximum $E'/E$ becomes
 \begin{equation}
   \left. \frac{E'}{E} \right|_{max} = \left| \frac{S'}{S} \right|
 \end{equation}
 along the direction $(p, q) = (0, 0)$, i.e. $\theta = 0$.  Since
numerical integrations will only be done along this path and the $\theta =
\pi$ one, we treat $E$ as a function of $r$ only.  We make $E'/E$ as large
as possible without violating the no shell crossings condition $E'/E \leq
M'/3M$ with
 \begin{equation}
    E = E_0 (1 + E_1 r^2) + E_2 ~~,~~~~~~~~ E_0, E_1, E_2 > 0,
 \end{equation}
 where the shell crossing occurs somewhere if $E_2 = 0$.

 Because of the two reflection symmetries, we can start integrating a null
ray from maximum expansion at the neck,
 \begin{equation}
   \eta = \pi ~~,~~~~~~~~ r = 0,
 \end{equation}
 where AH$^+$ \& AH$^-$ meet.  The symmetry means that integrating forwards
along increasing $r$ \& $t$ and integrating backwards along decreasing $r$
\& $t$ is the same thing, so one integration actually traces both halves
of the same ray.  Rays that don't pass through this point require two
separate parts to the integration, one from maximum expansion towards $r$
\& $t$ increasing, and the other towards $r$ \& $t$ decreasing, with
careful treatment of the neck limits where $r$ goes through zero.

\begin{figure}
 \begin{center}
 \parbox{14cm}{
 \includegraphics[scale = 0.85]{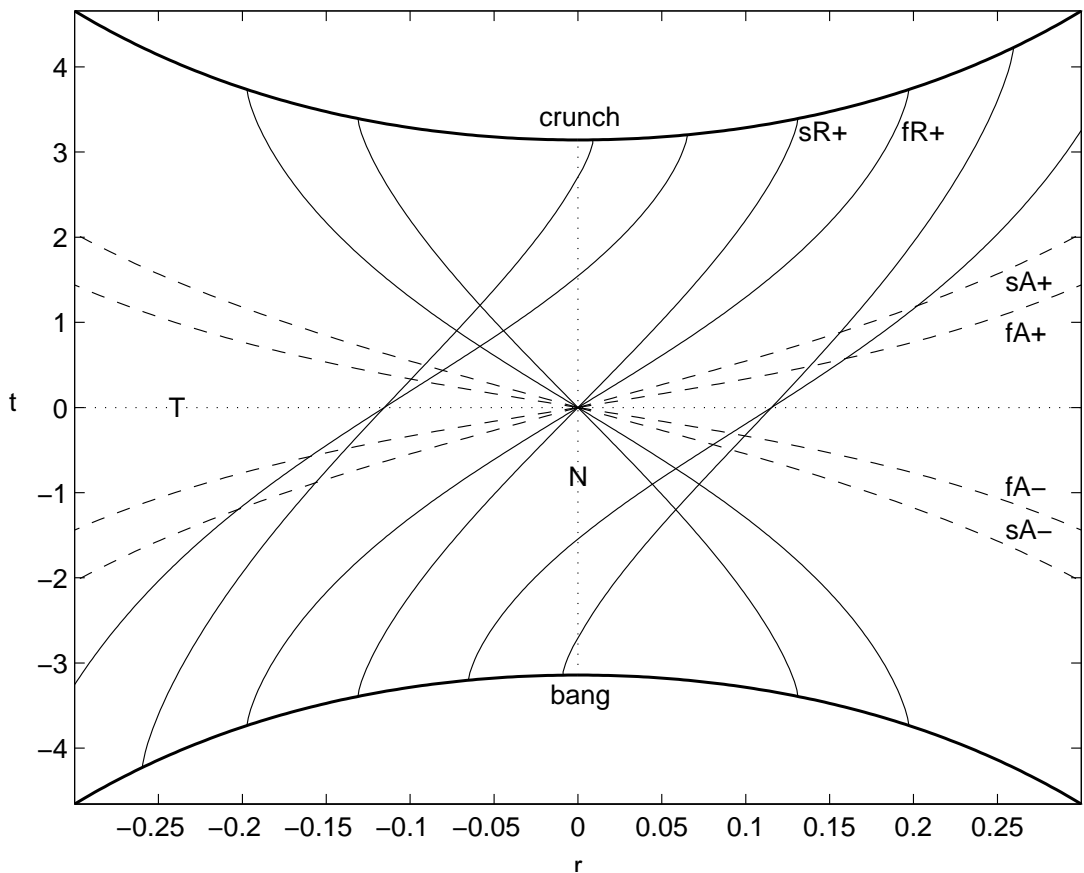}
 \caption{
 \label{SZrun1}
 \footnotesize
 The $(r-t)$ diagram for the Szekeres model defined for run 1, showing the
fast and slow future apparent horizons (fA+ and sA+), and past apparent
horizons, the fast and slow rays that pass through O
 --- the neck at the moment of maximum expansion
 --- towards $r$ increasing (fR+ and sR+), and rays through O going
towards $r$ decreasing, as well as rays going through other points.  T is
the moment of time symmetry which is also the simultaneous time of maximum
expansion, and N is the locus of the neck $r = 0$.  Note that fA+ \& sA+
are two different intersections of the future apparent horizon AH$^+$ in
two different radial directions
 --- the fast \& slow poles where $E'/E$ takes extreme values.  Note also
that there is no origin $R(r = r_o, t) = 0$ in wormhole models.}
 }
 \end{center}
 \end{figure}

 The following runs were done:
 \begin{itemize}
 \item   Test 1
 --- the vacuum case:
 \begin{equation}
   M_0 = 1~,~~ M_1 = 0~,~~ E_0 = 1~,~~ E_1 = 0~,~~ E_2 = 0~,~~ r_s = 1.
 \end{equation}
 As expected, we found that the fast AH, the slow AH, the fast null ray,
and the slow null ray were all the same.
 \item   Test 2
 --- the LT case:
 \begin{equation}
   M_0 = 1~,~~ M_1 = 0.1~,~~ E_0 = 1~,~~ E_1 = 0~,~~ E_2 = 0.01~,~~
   r_s = 1.
 \end{equation}
 Here the fast \& slow rays were the same, and the fast \& slow AHs were
the same, but the rays fell inside the AHs, as expected.
 \item   Test 3
 --- a Szekeres version of above LT case:
 \begin{equation}
   M_0 = 1~,~~ M_1 = 0.1~,~~ E_0 = 1~,~~ E_1 = 0.1~,~~ E_2 = 0.01~,~~
   r_s = 1.
 \end{equation}
 The AHs and rays were split on either side of the Test 2 curves.
 \item   Run 1
 --- medium  $M'/f'$
 \begin{equation}
   M_0 = 1~,~~ M_1 = 1~,~~ E_0 = 1~,~~ E_1 = 1~,~~ E_2 = 0~,~~
   r_s = 1.
 \end{equation}
 We found that the rays \& AHs were well split, while the rays were
strongly trapped.
 \item   Run 2
 --- low  $M'/f'$
 \begin{equation}
   M_0 = 1~,~~ M_1 = 1~,~~ E_0 = 1~,~~ E_1 = 1~,~~ E_2 = 0.1~,~~
   r_s = 0.01.
 \end{equation}
 The rays were mildly split, the AHs were indistinguishable in the range
plotted, and the rays were mildly trapped.
 \item   Run 3
 --- slightly less low $M'/f'$
 \begin{equation}
   M_0 = 1~,~~ M_1 = 2~,~~ E_0 = 1~,~~ E_1 = 2~,~~ E_2 = 0.01~,~~
   r_s = 0.01.
 \end{equation}
 This was very similar to the previous run, with the rays less mildly
trapped.
 \item   Run 4
 --- high  $M'/f'$
 \begin{equation}
   M_0 = 1~,~~ M_1 = 3~,~~ E_0 = 1~,~~ E_1 = 3~,~~ E_2 = 0.0001~,~~
   r_s = 10.
 \end{equation}
 Here the rays \& AHs were well split, and the rays were very strongly
trapped.
 \end{itemize}
These examples cover the main possibilites, and run 1 is shown in Fig.
\ref{SZrun1}.

 \section{Conclusions}

 Szekeres (S) models are a generalisation of the spherically symmetric
 Lema\^{\i}tre-Tolman (LT) models.  Both describe inhomogeneous dust
distributions, but the former have no Killing vectors.  There are 3
arbitrary functions of coordinate radius in LT models ($M$, $f$ \& $a$),
and a further 3 in S models ($S$, $P$ \& $Q$).

 For quasi-spherical Szekeres (S) models, we established 3 sets of
regularity conditions
 --- the conditions for a regular origin,
 the conditions for no shell crossings,
 and the conditions for regular maxima and minima in the spatial sections.
 The last two contain exactly those for the LT models, with extra
conditions on the arbitrary functions that are peculiar to S.  Thus, for
every regular LT model that is
 non-vacuum ($M' > 0$) at least somewhere, one can find regular S models
that are anisotropic versions of the same topology.  (For vacuum, $M' =
0$, S models must be spherically symmetric.)

 Since LT models can reproduce the
 Schwarzschild-Kruskal-Szekeres topology of a wormhole connecting two
universes, but with
 non-zero density everywhere, this is also possible with S models.  In the
vacuum case ($M' = 0$) this gives the full Kruskal manifold in geodesic
coordinates.  It is known that the presence of matter in such models inhibits
communication through the wormhole and splits the event horizons. We
investigated the S wormhole models, considering apparent horizons and the paths
of `radial' null rays, which, while not geodesic, are the fastest paths out of
a wormhole.  We showed that, even though the S model's anisotropy makes the
proper separation of consecutive shells shorter along certain directions, and
null motion faster along those same directions, this is not enough to
compensate for the retarding effect of matter.  Thus the causal structure of an
S wormhole is the same as that of the corresponding LT model.

 We also considered whether the two universes on either side of a wormhole
could be joined across a 3-surface, making a handle topology.  It was
found that a smooth junction is not possible at any finite distance, as a
surface layer would be created.  This conclusion applies to LT models and
to the vacuum case
 --- a Schwarzschild wormhole
 --- too.

 \setcounter{secnumdepth}{0}
 \section{Acknowledgments}
 \setcounter{secnumdepth}{1}

 A.K. is grateful to the University of Cape Town, where this work was started,
for hospitality, support and inspiring interactions.

\appendix

\section{The hypersurface of zero mass-dipole intersects every $(t = {\rm
const}, r = {\rm const})$ sphere.} \label{massdipole}

This hypersurface is given by

\begin{equation}\label{zerodipsurf}
E'/E = (A' + C')/(A + C).
\end{equation}
Since $E'/E$ at constant $r$ is bounded (see (\ref{E'Eextreme})), it must be
verified whether eq. (\ref{zerodipsurf}) has a solution in every sphere of
constant $t$ and $r$. The solution will exist when

\begin{equation}\label{condzerodip}
(E'/E)_{\rm min} \leq (A' + C')/(A + C) \leq (E'/E)_{\rm max}.
\end{equation}
Since $(E'/E)_{\rm min} = - (E'/E)_{\rm max}$, eq. (\ref{condzerodip}) is
equivalent to

\begin{equation}\label{condzerodip2}
(A' + C')^2/(A + C)^2 \leq (E'/E)^2_{\rm extreme} = \frac 1 {S^2}\left({P'}^2 +
{Q'}^2 + {S'}^2\right).
\end{equation}
We have
\begin{equation}\label{ACvsPQS}
A + C = \frac 1{2S}\left(1 + P^2 + Q^2 + S^2\right)
 $$ $$
 A' + C' = \frac
{S'}{2S^2}\left(S^2 - P^2 - Q^2 - 1\right) + \frac 1S\left(PP' + QQ'\right).
\end{equation}
Substituted in (\ref{condzerodip2}), this leads to

\begin{equation}\label{ineqforS'}
4S^2{S'}^2\left(1 + P^2 + Q^2\right) - 4SS'\left(PP' + QQ'\right)\left(S^2 -
P^2 - Q^2 - 1\right)
 $$ $$
 - 4S^2(PP' + QQ')^2 + \left({P'}^2 + {Q'}^2\right)
\left(1 + P^2 + Q^2 + S^2\right) \geq 0.
\end{equation}
The discriminant of this with respect to $S'$ is

\begin{equation}\label{discrS'}
\Delta = -16S^2\left(1 + P^2 + Q^2 + S^2\right)^2\left[\left(PQ' - QP'\right)^2
+ {P'}^2 + {Q'}^2\right],
\end{equation}
and is always negative unless $P' = Q' = 0$. This means that with $(P', Q')
\neq (0, 0)$, the l.h.s. of (\ref{ineqforS'}) is strictly positive. Even when
$P' = Q' = 0$, it is still strictly positive unless $S' = 0$ as well. However,
$P' = Q' = S' = 0$ implies $A' = C' = 0$ and $E' = 0$ on the whole sphere, and
then the dipole component of density $\Delta \rho = 0$; i.e. on such a sphere
the density is spherically symmetric. Hence, apart from the spherically
symmetric subcase, eq. (\ref{condzerodip}) is fulfilled, with sharp
inequalities in both places. This means that the $\Delta\rho = 0$ hypersurface
intersects every $(t = {\rm const}, r = {\rm const})$ sphere along a circle
parallel to the $E' = 0$ circle (see remark after eq. (\ref{E'/Econst})).

 \section{Matching the Szekeres Metric to Itself}
 \label{JCcalcs}

 We here lay out the calculations necessary for matching the Szekeres
metric across a comoving surface to some other metric, and in particular
to another Szekeres metric.

 Given a comoving surface,
 \begin{equation}
   r_\Sigma = Z(p,q)
 \end{equation}
 and surface coordinates,
 \begin{equation}
   \xi^i = (t, p, q)
 \end{equation}
 we calculate the basis vectors in the surface,
 \begin{equation}
   e_i^\mu = \frac{\partial x^\mu_\pm}{\partial \xi^i},
 \end{equation}
 the 1st fundamental form,
 \begin{equation}
   {}^3\!\!g_{ij}^+ = {}^3\!\!g_{ij}^- = g_{\mu\nu}^\pm e_i^\mu e_j^\nu,
 \end{equation}
 the normal vector,
 \begin{equation}
   n_\mu ~~,~~~~~~~~ n_\mu n^\mu = 1 ~~,~~~~~~~~ n_\mu e_i^\mu = 0,
 \end{equation}
 and the 2nd fundamental form
 \begin{equation}
   K^\pm_{ij} = - n^\pm_\lambda \left(
      \frac{\partial^2 x^\lambda}{\partial \xi^i \partial \xi^j}
      + \Gamma^\lambda_{\mu\nu} \frac{\partial x^\mu}{\partial \xi^i}
      \frac{\partial x^\nu}{\partial \xi^j}  \right).
 \end{equation}

 Using GRTensor/GRJunction \cite{MPL96}
 we find the following for the intrinsic metric:
 \begin{eqnarray}
   {}^3\!\!g_{tt} & = &  - 1, \\ \nonumber \\
   {}^3\!\!g_{pp} & = & \frac{Z_p^2 (R' - RE'/E)^2 E^2 + R^2 (\epsilon + f)}
      {E^2 (\epsilon + f)}, \\ \nonumber \\
   {}^3\!\!g_{pq} & = & \frac{Z_p Z_q (R' - RE'/E)^2}{(\epsilon + f)}, \\
\nonumber \\
   {}^3\!\!g_{qq} & = & \frac{Z_q^2 (R' - RE'/E)^2 E^2 + R^2 (\epsilon + f)}
      {E^2 (\epsilon + f)},
 \end{eqnarray}
 the surface basis vectors:
 \begin{eqnarray}
   e^t_i & = & (1, 0, 0), \\
   e^r_i & = & (0, Z_p, Z_q), \\
   e^p_i & = & (0, 1, 0), \\
   e^q_i & = & (0, 0, 1),
 \end{eqnarray}
 the surface normal:
 \begin{eqnarray}
   n_r & = & - \frac{(R' E - R E') R}{E \Delta}, \\ \nonumber \\
   n_p & = & \frac{Z_p (R' E - R E') R}{E \Delta}, \\ \nonumber \\
   n_q & = & \frac{Z_q (R' E - R E') R}{E \Delta}, \\ \nonumber \\
 \mbox{where~~~~} &&
      Z_p = \frac{\partial Z}{\partial p} ~~,~~~~~~~~
      Z_q = \frac{\partial Z}{\partial q} ~~, \\ \nonumber \\
   \Delta & = & \left( R^2 (\epsilon + f) +
      (Z_p^2 + Z_q^2) (R' E - R E')^2 \right)^{1/2},
 \end{eqnarray}
 and the extrinsic curvature:
 \begin{eqnarray}
   K_{pt} & = & \frac{[ Z_p ( R \dot{R}' - \dot{R} R' ) ]}{\Delta},
      \label{Kpt} \\ \nonumber \\
   K_{qt} & = & \frac{[ Z_q ( R \dot{R}' - \dot{R} R') ]}{\Delta},
      \label{Kqt} \\ \nonumber \\
   K_{pp} & = & \frac{1}{2 E^2 (\epsilon + f) \Delta}
      \Bigg\{ 2 R E (R' E - R E') (\epsilon + f) Z_{pp}
      \nonumber \\ \nonumber \\
   && + 2 (R' E - R E')^2 (E_p E' - E'_p E) Z_p^3 \nonumber \\
   && + 2 (R' E - R E')^2 (E_q E' - E'_q E) Z_p^2 Z_q \nonumber \\
   && + \Big\{ 2 \left[ 3 E R' R E' + R R'' E^2 - (E')^2 R^2 - 2 E^2 (R')^2
      - R^2 E'' E \right] (\epsilon + f) \nonumber \\
   &&~~~ - R  f' E (R' E - R E') \Big\} Z_p^2 \nonumber \\
   && - 2 R (2 R E'_p E - E_p R' E -R E_p E') (\epsilon + f) Z_p
      \nonumber \\
   && - 2 R E_q (R' E - R E') (\epsilon + f) Z_q \nonumber \\
   && - 2 R^2 (\epsilon + f)^2 \Bigg\}, \label{Kpp} \\ \nonumber \\
   K_{pq} & = & \frac{1}{2 E^2 (\epsilon + f) \Delta}
      \Bigg\{ 2 R E (R' E - R E') (\epsilon + f) Z_{pq} \nonumber \\
   && + 2 (R' E - R E')^2 (E_p E' - E'_p E) Z_p^2 Z_q \nonumber \\
   && + 2 (R' E - R E')^2 (E_q E' - E'_q E) Z_p Z_q^2 \nonumber \\
   && + \Big\{ 2 \left[ 3 E R' R E' + R R'' E^2 - (E')^2 R^2 - 2 E^2 (R')^2
      - R^2 E'' E \right] (\epsilon + f) \nonumber \\
   &&~~~ - R f' E (R' E - R E') \Big\} Z_p Z_q \nonumber \\
   && - 2 R E (R E'_q - R' E_q) (\epsilon + f) Z_p \nonumber \\
   && - 2 R E (R E'_p - R' E_p) (\epsilon + f) Z_q \Bigg\} ,
      \label{Kpq} \\ \nonumber \\
   K_{qq} & = & \frac{1}{2 E^2 (\epsilon + f) \Delta}
      \Bigg\{  2 R E (R' E - R E') (\epsilon + f) Z_{qq} \nonumber \\
   && + 2 (R' E - R E')^2 (E_p E' - E'_p E) Z_p Z_q^2 \nonumber \\
   && + 2 (R' E - R E')^2 (E_q E' - E'_q E) Z_q^3 \nonumber \\
   && + \Big\{ 2 \left[ 3 E R' R E' + R R'' E^2 - (E')^2 R^2 - 2 E^2 (R')^2
      - R^2 E'' E \right] (\epsilon + f) \nonumber \\
   &&~~~ - R f' E (R' E - R E') \Big\} Z_q^2 \nonumber \\
   && - 2 R E_p (R' E - R E') (\epsilon + f) Z_p \nonumber \\
   && - 2 R (2 R E'_q E - E_q R' E - R E_q E') (\epsilon + f) Z_q
      \nonumber \\
   && - 2 R^2 (\epsilon + f)^2 \Bigg\} , \label{Kqq}
 \end{eqnarray}
 where all quantities are evaluated on $\Sigma$.

 \section{The Acceleration of a Given Tangent Vector}
 \label{AccelCalcs}

 Starting from
 \begin{equation}
   a^\alpha = k^\beta \nabla_\beta k^\alpha =
      k^\beta \partial_\beta k^\alpha
      + \Gamma^\alpha{}_{\beta\gamma} k^\beta k^\gamma, \\
 \end{equation}
 the individual acceleration components for a given $k^\alpha$ in the
Szekeres metric are
 \begin{eqnarray}
   a^t & = & k^\beta \partial_\beta k^t
      + \Gamma^t{}_{rr} (k^r)^2
      + \Gamma^t{}_{pp} (k^p)^2
      + \Gamma^t{}_{qq} (k^q)^2 \\
   & = & k^t \partial_t k^t + k^r \partial_r k^t + k^p \partial_p k^t
         + k^q \partial_q k^t   \nonumber \\ \nonumber \\
   && + \left( R' - \frac{R E'}{E} \right) \left( \dot{R}' -
         \frac{\dot{R} E'}{E} \right) \frac{1}{\epsilon + f} (k^r)^2
      + \left( \frac{R \dot{R}}{E^2} \right) \left( (k^p)^2 + (k^q)^2
         \right), \\ \nonumber \\
   a^r & = & k^\beta \partial_\beta k^r
      + 2 \Gamma^r{}_{tr} k^t k^r
      + \Gamma^r{}_{rr} (k^r)^2
      + 2 \Gamma^r{}_{rp} k^r k^p
      + 2 \Gamma^r{}_{rq} k^r k^q   \nonumber \\
      && + \Gamma^r{}_{pp} (k^p)^2
      + \Gamma^r{}_{qq} (k^q)^2 \\ \nonumber \\
   & = & k^t \partial_t k^r + k^r \partial_r k^r + k^p \partial_p k^r
         + k^q \partial_q k^r
      + \frac{2 \left( \dot{R}' - \frac{\dot{R} E'}{E} \right)}
         {\left( R' - \frac{R E'}{E} \right)} k^t k^r
      \nonumber \\ \nonumber \\
   && + \left[ \frac{\left( R' - \frac{R E'}{E} \right)'}
         {\left( R' - \frac{R E'}{E} \right)}
         - \frac{f'}{2(\epsilon + f)} \right] (k^r)^2
      - \frac{2 R \left( \frac{E'_p}{E} - \frac{E' E_p}{E^2} \right)}
         {\left( R' - \frac{R E'}{E} \right)} k^r k^p
      \nonumber \\ \nonumber \\
   && - \frac{2 R \left( \frac{E'_q}{E} - \frac{E' E_q}{E^2} \right)}
         {\left( R' - \frac{R E'}{E} \right)} k^r k^q
      - \frac{R (\epsilon + f)}{E^2 \left( R' - \frac{R E'}{E} \right)}
         \left( (k^p)^2 + (k^q)^2 \right), \\ \nonumber \\
   a^p & = & k^\beta \partial_\beta k^p
      + 2 \Gamma^p{}_{tp} k^t k^p
      + \Gamma^p{}_{rr} (k^r)^2
      + 2 \Gamma^p{}_{rp} k^r k^p   \nonumber \\
      && + \Gamma^p{}_{pp} (k^p)^2
      + 2 \Gamma^p{}_{pq} k^p k^q
      + \Gamma^p{}_{qq} (k^q)^2 \\ \nonumber \\
   & = & k^t \partial_t k^p + k^r \partial_r k^p + k^p \partial_p k^p
         + k^q \partial_q k^p
      + \frac{2 \dot{R}}{R} k^t k^p   \nonumber \\ \nonumber \\
   && - \frac{\left( R' - \frac{R E'}{E} \right) (E E'_p - E' E_p)}
         {R (\epsilon + f)} (k^r)^2
      + \frac{2 \left( R' - \frac{R E'}{E} \right)}{R} k^r k^p
            \nonumber \\ \nonumber \\
   && - \frac{E_p}{E} (k^p)^2
      - \frac{2 E_q}{E} k^p k^q
      + \frac{E_p}{E} (k^q)^2, \\ \nonumber \\
   a^q & = & k^\beta \partial_\beta k^q
      + 2 \Gamma^q{}_{tq} k^t k^q
      + \Gamma^q{}_{rr} (k^r)^2
      + 2 \Gamma^q{}_{rq} k^r k^q   \nonumber \\
      && + \Gamma^q{}_{pp} (k^p)^2
      + 2 \Gamma^q{}_{pq} k^p k^q
      + \Gamma^q{}_{qq} (k^q)^2 \\ \nonumber \\
   & = & k^t \partial_t k^q + k^r \partial_r k^q + k^p \partial_p k^q
         + k^q \partial_q k^q
      + \frac{2 \dot{R}}{R} k^t k^q   \nonumber \\ \nonumber \\
   && - \frac{\left( R' - \frac{R E'}{E} \right) (E E'_q - E' E_q)}
         {R (\epsilon + f)} (k^r)^2
      + \frac{2 \left( R' - \frac{R E'}{E} \right)}{R} k^r k^q
            \nonumber \\ \nonumber \\
   && + \frac{E_q}{E} (k^p)^2
      - \frac{2 E_p}{E} k^p k^q
      - \frac{E_q}{E} (k^q)^2.
 \end{eqnarray}
 For ``radial" paths $k^p = 0 = k^q$, $\partial_p k^\alpha = 0 =
\partial_q k^\alpha$ these reduce to
 \begin{eqnarray}
   a^t & = & k^t \partial_t k^t + k^r \partial_r k^t
   \nonumber \\ \nonumber \\
   && + \left( R' - \frac{R E'}{E} \right) \left( \dot{R}' -
         \frac{\dot{R} E'}{E} \right) \frac{1}{(\epsilon + f)} (k^r)^2,
   \\ \nonumber \\
   a^r & = & k^t \partial_t k^r + k^r \partial_r k^r
      + \frac{2 \left( \dot{R}' - \frac{\dot{R} E'}{E} \right)}
         {\left( R' - \frac{R E'}{E} \right)} k^t k^r
   \nonumber \\ \nonumber \\
   && + \left[ \frac{\left( R' - \frac{R E'}{E} \right)'}
         {\left( R' - \frac{R E'}{E} \right)}
         - \frac{f'}{2(\epsilon + f)} \right] (k^r)^2,
   \\ \nonumber \\
   a^p & = & - \frac{\left( R' - \frac{R E'}{E} \right) (E E'_p - E' E_p)}
         {R (\epsilon + f)} (k^r)^2,
   \\ \nonumber \\
   a^q & = & - \frac{\left( R' - \frac{R E'}{E} \right) (E E'_q - E' E_q)}
         {R (\epsilon + f)} (k^r)^2.
 \end{eqnarray}
 Using the ``radial" null condition
 \begin{equation}
   k^t = \frac{j k^r}{\sqrt{\epsilon + f}\;}
      \left( R' - \frac{R E'}{E} \right)
 \end{equation}
 the acceleration becomes
 \begin{eqnarray}
   a^t & = & \frac{j}{\sqrt{\epsilon + f}\;} \left\{
      \left( \dot{R}' - \frac{\dot{R} E'}{E} \right) k^t k^r
      + \left( R' - \frac{R E'}{E} \right) k^t \partial_t k^r \right\}
      \nonumber \\ \nonumber \\
      && + \frac{j}{\sqrt{\epsilon + f}\;} \left\{
      \left[ \left( R' - \frac{R E'}{E} \right)'
      - \frac{f'}{2 (\epsilon + f)} \left( R' - \frac{R E'}{E} \right)
      \right] (k^r)^2   \right.   \nonumber \\ \nonumber \\
      &&   \left.
      + \left( R' - \frac{R E'}{E} \right) k^r \partial_r k^r \right\}
      + \left( R' - \frac{R E'}{E} \right) \left( \dot{R}' -
         \frac{\dot{R} E'}{E} \right) \frac{1}{(\epsilon + f)} (k^r)^2,
      \\ \nonumber \\
   a^r & = & k^r \partial_r k^r
      + \frac{j}{\sqrt{\epsilon + f}\;} \left\{
      \left( R' - \frac{R E'}{E} \right) k^r \partial_t k^r
      + 2 \left( \dot{R}' - \frac{\dot{R} E'}{E} \right)
      (k^r)^2   \right\}   \nonumber \\ \nonumber \\
   && + \left[ \frac{\left( R' - \frac{R E'}{E} \right)'}
         {\left( R' - \frac{R E'}{E} \right)}
         - \frac{f'}{2(\epsilon + f)} \right] (k^r)^2, \\ \nonumber \\
   a^p & = & - \frac{\left( R' - \frac{R E'}{E} \right) (E E'_p - E' E_p)}
         {R (\epsilon + f)} (k^r)^2, \\ \nonumber \\
   a^q & = & - \frac{\left( R' - \frac{R E'}{E} \right) (E E'_q - E' E_q)}
         {R (\epsilon + f)} (k^r)^2.
 \end{eqnarray}
 While $E' = 0$ gives the expected LT values, we note that $E(r, p, q)$
determines whether $a^p$ and $a^q$ are zero or not.

 In the $\epsilon = +1$ case, by (\ref{Riemprojp}), (\ref{TanPhiX}) and
(\ref{TanThetaX}), the extremes of $E$ on a given
 2-sphere are located at
 \begin{eqnarray}
   p_e & = & P + \frac{P' S}
      {\Big( \pm \sqrt{(S')^2 + (P')^2 + (Q')^2}\; - S' \Big)} \\
   q_e & = & Q + \frac{Q' S}
      {\Big( \pm \sqrt{(S')^2 + (P')^2 + (Q')^2}\; - S' \Big)}
 \end{eqnarray}
 and it is easily verified that $a^p = 0 = a^q$ in these two antipodal
directions.  It follows that initially radial geodesics in these
directions remain radial if $p_e$ \& $q_e$ are constant with $r$.  For
example, if $p_e = 0 = q_e$, this would require arbitrary functions
satisfying
 \begin{eqnarray}
   && \frac{P'}{P} = \frac{2 S' S}{S^2 - P^2 - Q^2} = \frac{Q'}{Q} \\
   \mbox{or}~~~~ &&
      P = 0 ~~,~~~~~~ \frac{Q'}{Q} = \frac{2 S' S}{S^2 - Q^2} ~~,
      ~~~~\mbox{when}~~ P' = 0 \\
   \mbox{or}~~~~ &&
      \frac{P'}{P} = \frac{2 S' S}{S^2 - P^2} ~~,~~~~~~ Q = 0 ~~,
      ~~~~\mbox{when}~~ Q' = 0 \\
   \mbox{or}~~~~ &&
      P = 0 ~~,~~~~~~ Q = 0 ~~, ~~~~\mbox{when}~~ P' = 0
      ~~~~\mbox{and}~~~~ Q' = 0
 \end{eqnarray}

 \section{Other Features of the AH}
 \label{AHfeatures}

 \subsection{The FLRW Case}

 The dust FLRW limit is $M = M_0 r^3$, $f = - k r^2$, $a = 0$, $R = r
S(t)$, and the $R = 2M$ locus is given by
 \begin{equation}
   S(t_{AH}) = 2 M_0 r^2
 \end{equation}
 In the collapse phase of a $k = +1$ model, the time of the future AH is
 \begin{eqnarray}
   \cos \eta_{AH} & = & 1 - 2 r^2 \\
   \rightarrow~~~~ t_{AH} & = &
      M_0 \left[ \pi + \arccos(2 r^2 - 1) + 2 r \sqrt{1 - r^2}\; \right]
 \end{eqnarray}
 which has slope
 \begin{equation}
   \left( \frac{dt}{dr} \right)_{AH} =
      - \frac{4 M_0 r^2}{\sqrt{1 - r^2}\;}
 \end{equation}
 while the light rays have slopes
 \begin{equation}
   \left( \frac{dt}{dr} \right)_n =
      \pm \frac{S}{\sqrt{1 - r^2}\;} =
      \pm \frac{2 M_0 r^2}{\sqrt{1 - r^2}\;}
 \end{equation}
 Clearly the future AH is incoming timelike.  The result extends to all
$k$ values, and the converse holds for the past AH in the expansion phase.

 \subsection{Behaviour Near the Bang or Crunch}

 Consider eq (\ref{etaAHeq}) for the locus of the future AH in a
collapsing elliptic region, $f < 0$, $\pi < \eta \leq 2 \pi$, in terms of
parameter $\eta$.  Near the crunch, $\overline{\eta} = 2 \pi - \eta
\rightarrow 0$, we find
 \begin{eqnarray}
   0 & = &
      \left[
      1 - \sqrt{\frac{(-f)}{(1 + f)}}\; \left( \frac{2}{\overline{\eta}}
      \right) \right] \left\{
      \frac{(-f)^{3/2} a'}{M} \left( \frac{4}{\overline{\eta}^3} \right)
      - \frac{f'}{f} \left( \frac{12 \pi}{\overline{\eta}^3} \right)
      + \frac{M'}{M} \left( \frac{8 \pi}{\overline{\eta}^3} \right)
      \right\} \nonumber \\
   && + \left[
      \sqrt{\frac{(-f)}{(1 + f)}}\; \left( \frac{2}{\overline{\eta}}
      \right) \right] \frac{E'}{E}.
 \end{eqnarray}
 As noted previously, when $E' = 0$, the solution makes the first bracket
zero, $\overline{\eta} \approx 2 \sqrt{-f}\; \rightarrow 0$.  (Even in this
case, where we know $R = 2M$ is the AH, the fact that $R'$ diverges at $R = 0$
means we must multiply through by $\overline{\eta}$ to make the rhs zero
there.)  Notice too that the no shell crossing condition (\ref{crunchCondE})
ensures the second bracket is generically
 non-negative where $M' > 0$ and
 non-positive where $M' < 0$.  Assuming we aren't near an origin, $0 < M <
\infty$, it is clear that, even if $f \rightarrow 0$, the last two terms
in this second bracket are divergent, with the middle one dominant, making
the $E'$ term negligible.  Thus $\overline{\eta} \approx 2 \sqrt{-f}\;$ is
still the solution in the limit.  However (\ref{ellevtS}) in the
$\overline{\eta} \approx 2 \sqrt{-f}\; \rightarrow 0$ limit shows that the
time from AH to crunch goes to
 \begin{equation}
   b - t_{AH} \rightarrow \frac{M}{(-f)^{3/2}} \frac{\overline{\eta}^3}{6}
   \rightarrow \frac{4 M}{3}
 \end{equation}
 where the crunch time $b(r)$ is defined in (\ref{Def_b(r)}).  Therefore
the future AH does not intersect the crunch away from an origin.  The
result is just the time reverse for the past AH near the bang in an
expanding region, and a similar calculation applies for hyperbolic or
extended parabolic regions, giving the same result.

 An $f = 0$ locus is where an interior elliptic region joins to an
exterior hyperbolic or parabolic region.  The transition involves the
lifetime of the worldlines diverging, so either the crunch goes to the
infinite future, or the bang goes to the infinite past.  Since there is
only one AH in a hyperbolic or parabolic region, one of the two AH loci in
the elliptic region also exits to infinity before $f = 0$ is reached.

 A third possibility where $f$ is only asymptotically zero (the
asymptotically flat case) is that both the bang and the crunch diverge to
the infinite past and future, and the two AHs go with them.

 \subsection{Behaviour Near an Origin}

 Consider eq (\ref{etaAHeq}) again.  Near a regular origin, along a
constant $t$ or constant $\eta$ surface (see section \ref{RegOrig}), $M
\approx \mu (-f)^{3/2}$, $E \approx \nu (-f)^{n/2}$ for some positive
constants $\mu$ \& $\nu$, and $f \rightarrow 0$, so that
 \begin{eqnarray}
   0 & = & \left[ 1 + \sqrt{-f}\; \frac{\sin \eta}{(1 - \cos \eta)}
      \right] \Bigg\{
      - \frac{a'}{\mu} \left( \frac{\sin \eta}{(1 - \cos \eta)^2} \right)
      + \frac{f'}{2f} \Bigg\}
      \nonumber \\
   && - \left[ \sqrt{-f}\; \frac{\sin \eta}{(1 - \cos \eta)}
      \right] \frac{nf'}{2f}.
 \end{eqnarray}
 We divide through by $f'$ and define $X = - \sqrt{-f}\; \frac{\sin
\eta}{(1 - \cos \eta)}$ which is positive for $\eta > \pi$, giving
 \begin{eqnarray}
   0 & = & [1 - X] \Bigg\{
      - \frac{a'}{\mu f'} \left( \frac{\sin \eta}{(1 - \cos \eta)^2} \right)
      + \frac{1}{2f}
      + \frac{X}{1 - X} \frac{n}{2f} \Bigg\}.
 \end{eqnarray}
 Though $a'$ and $M'$ always have opposite signs, $f'$ may have either
sign in an elliptic region, but in general we don't expect terms to cancel
in the curly brackets.  Thus, whether or not the $a'$ term diverges, we
must have $X \rightarrow 1$, i.e. $\sin \eta \rightarrow 0$ so that
 \begin{equation}
   \overline{\eta} = 2 \pi - \eta \rightarrow 2 \sqrt{-f}.
 \end{equation}
 Unlike the previous case, though, $M \rightarrow 0$ ensures the AH
intersects the crunch here,
 \begin{equation}
   b - t_{AH} \rightarrow \mu \frac{\overline{\eta}^3}{6} \rightarrow 0.
 \end{equation}

 As always, the time reverse applies in an expanding phase, and the
hyperbolic and parabolic cases give the same result.

 \end{document}